%% file: main.tex
\documentclass[
  acmsmall,
  screen,
  nonacm
]{acmart}
\usepackage{amsmath}
\usepackage{mathtools}
\usepackage{booktabs}
\usepackage{xspace}
\usepackage{microtype}
\usepackage{enumitem}
\usepackage{listings}
\usepackage{tikz}
\usetikzlibrary{positioning,arrows.meta}
\usepackage{stmaryrd}
\usepackage{mathpartir}
\usepackage{multirow}
\usepackage[capitalise,nameinlink]{cleveref}
\AtBeginEnvironment{definition}{\crefalias{theorem}{definition}}
\AtBeginEnvironment{lemma}{\crefalias{theorem}{lemma}}

\setcopyright{none}

\AtEndPreamble{\counterwithout{theorem}{section}}

\newcommand{\C}[1]{{\small\texttt{#1}}}
\newcommand{\Cs}[1]{{\scriptsize\texttt{#1}}}

\newcommand{\mcomment}[1]{\text{\color{lstcomment}// #1}}
\newcommand{\sem}[1]{\llbracket #1 \rrbracket}
\newcommand{\pf}[1]{\textcolor{lstfunc}{#1}}
\newcommand{\imop}[1]{\textcolor{lstimag}{\mathsf{#1}}}

\newcommand{\pure}[1]{\ulcorner #1 \urcorner}
\newcommand{\reg}[1]{\mathtt{\%#1}}
\newcommand{\Arr}[1]{\mathsf{Array}\ \mathtt{#1}}
\newcommand{\Vect}[1]{\mathsf{Vec}\ \mathtt{#1}}
\newcommand{\rep}[4][]{\mathsf{own}_{#1}\, #2 \triangleleft #3\ #4}
\newcommand{\pair}[5][]{\mathsf{pair}_{#1}(#2, #3 \mid #4, #5)}
\newcommand{\bigsep}{\mathop{\scalebox{1.4}{$\ast$}}}
\newcommand{\toolname}{R2C2}

\DeclareCaptionSubType{figure}
\renewcommand{\thesubfigure}{(\alph{subfigure})}

\newcommand{\Wfun}{\mathsf{W}_{\mathsf{fun}}}
\newcommand{\Wcall}{\mathsf{W}_{\mathsf{call}}}
\newcommand{\Hide}{\mathsf{inline}}
\newcommand{\ctxref}{\sqsubseteq_{\mathrm{ctx}}}
\newcommand{\behref}{\sqsubseteq_{\mathrm{beh}}}

\newcommand{\res}[1]{\textcolor{lsttype}{#1}}
\newcommand{\cnum}[1]{\textbf{\textup{(#1)}}}
\newcommand{\irsk}[1]{\textcolor{lstkeyword}{\Cs{#1}}}
\newcommand{\irst}[1]{\textcolor{lsttype}{\Cs{#1}}}
\newcommand{\irsf}[1]{\textcolor{lstfunc}{\Cs{#1}}}

\newcommand{\dq}{q}

\definecolor{lstkeyword}{HTML}{046404}
\definecolor{lsttype}{HTML}{A3002B}
\definecolor{lstfunc}{HTML}{2060C0}
\definecolor{lstcomment}{HTML}{046404}
\definecolor{lstimag}{HTML}{A3002B}

\lstdefinelanguage{CPaper}[ANSI]{C}{
  deletekeywords={int,bool,char,float,double,void,long,short,unsigned,signed,const},
  morekeywords=[2]{int,bool,char,float,double,void,long,short,unsigned,signed,const,
    size_t, FILE, qoi_desc},
}

\lstdefinelanguage{LLVM}{
  morekeywords=[1]{define,declare,call,ret,br,alloca,load,store,getelementptr,
    add,sub,mul,shl,lshr,ashr,and,or,xor,icmp,fcmp,fadd,fsub,fmul,fdiv,
    phi,select,unreachable,zext,sext,trunc},
  morekeywords=[2]{void,ptr,label,i1,i8,i16,i32,i64,float,double},
  alsoletter={.\%@},
  morecomment=[l]{;},
  morestring=[b]",
  sensitive=true,
}

\lstdefinelanguage{Rust}{
  morekeywords=[1]{fn,let,mut,pub,return,if,else,for,in,while,loop,match,as,
    use,struct,impl,unsafe},
  morekeywords=[2]{i8,i16,i32,i64,u8,u16,u32,u64,usize,isize,f32,f64,bool,char,
    str,String,Vec,Option,Self, File, QoiDesc},
  morecomment=[l]{//},
  morecomment=[s]{/*}{*/},
  morestring=[b]",
  sensitive=true,
}

\lstdefinestyle{c}{language=CPaper}
\lstdefinestyle{rust}{language=Rust}
\lstdefinestyle{llvm}{language=LLVM}

\crefformat{section}{#2\S{}#1#3}

\begin{document}

\title{C-to-Rust Translation Validation at the LLVM IR Level via AI-Assisted Refinement Proof}

\author{Eunsu Kim}
\affiliation{%
  \institution{KAIST}
  \city{Daejeon}
  \country{Republic of Korea}
}
\email{es4761@kaist.ac.kr}
\author{Sukyoung Ryu}
\affiliation{%
  \institution{KAIST}
  \city{Daejeon}
  \country{Republic of Korea}
}
\email{sryu.cs@kaist.ac.kr}
\author{Jaemin Hong}
\affiliation{%
  \institution{UNIST}
  \city{Ulsan}
  \country{Republic of Korea}
}
\email{jaemin.hong@unist.ac.kr}

\input{sections/0-abstract}

\maketitle


\input{sections/1-intro}
\input{sections/2-1-llvm}
\input{sections/2-2-refinement}

\input{sections/3-validation}

\input{sections/4-common}
\input{sections/5-1-chain}
\input{sections/5-2-simulation}
\input{sections/6-automation}
\input{sections/7-1-eval}

\input{sections/7-2-eval}

\input{sections/8-related}

\input{sections/9-conclusion}

%


\bibliographystyle{ACM-Reference-Format}
\bibliography{reference}

\appendix
\input{sections/A-semantics}

\input{sections/B-compilation}

\end{document}

%% file: sections/0-abstract.tex
\begin{abstract}
Migrating C codebases to Rust is a promising approach
to securing legacy software, but existing translators do not guarantee that
a translated Rust program preserves the behavior of the C program.
Neither rule-based nor large language model (LLM)-based translators have been formally verified,
and existing approaches to checking translation correctness
are limited to testing and bounded checks,
which cannot reason about all possible executions.
In this work, we present \toolname{}, which validates C-to-Rust translation
by proving in Rocq that the Rust program \emph{refines} the C program, i.e.,
every behavior on the Rust side is also a behavior on the C side.
Since no formal semantics defines the behaviors of C and Rust programs in the same terms,
we prove refinement between the two programs compiled
to the LLVM intermediate representation (LLVM IR).
In addition, to resolve the discrepancies between the data representations of C and Rust,
we introduce a \emph{common abstraction} that relates two corresponding
representations through an abstract value, and specify
the behavior of each library function over abstract values.
We construct the refinement proof as a chain of refinement relations,
where only one relation is program-specific and proved at validation time,
while the others are proved for all programs in advance.
Proof construction is automated by AI agents,
and since Rocq checks the proofs, the agents need not be trusted.
We evaluate \toolname{} on LLM translations of 519 C programs drawn from a
benchmark suite used in prior work on C-to-Rust translation validation.
\toolname{} proves the translations of 486 programs correct
without bounding their executions,
and discovers three incorrect translations, providing valid counterexamples that
existing translation validation tools cannot find.
We also conduct a case study on a small real-world image compression
library and demonstrate how \toolname{} can handle file operations.
\end{abstract}

%% file: sections/1-intro.tex
\section{Introduction}

Migrating C codebases to memory-safe Rust is a promising approach to securing
legacy C software.
C often suffers from memory bugs because it lacks language-level protections,
whereas Rust provides memory safety guarantees at compile
time~\cite{rustbelt18}.
As a result, C-to-Rust translation has emerged as an active research area with a
variety of approaches, ranging from rule-based
translators~\cite{emre2021translating, emre2023aliasing, crown23,
hong2023concrat, hong2024dont, hong2024tag, wu2025genc2rust, hong2025forcrat,
fromherz2026scylla, kim2026securing, hong2026replacing, chen2026refinator, proctor26} to large
language model (LLM)-based methods~\cite{hong2024type, zhou2025c2rusttv,
xu2025optimizing, zhang2025systematic, vert25, rustassure25, cai2026rustmap,
shiraishi2026smartc2rust, yuan2026project}.

%

Unfortunately, existing translators do not guarantee the \emph{correctness} of
the translation, i.e., that a translated Rust program preserves the behavior of
the C program.
In theory, the correctness of rule-based translators could be proved, but no
existing translators have been formally verified.
LLM-based translators do not rely on algorithmic procedures that can be formally
verified.


When translators are not verified in advance, an alternative approach to
ensuring correctness is \emph{translation validation}~\cite{pnueli98}, which
instead verifies each individual translation.
While most prior studies adopt only testing, RustAssure~\cite{rustassure25}
conducts symbolic testing to expand the covered input space, and
VERT~\cite{vert25} leverages bounded model checking~\cite{cbmc04, kani22} to
compare behavior up to a fixed unrolling bound and recursion depth.
These approaches can discover incorrect translations but cannot reason about all
possible executions, leaving a significant gap in ensuring correctness.


To guarantee the correctness of translation for every possible execution, a line
of research on translation validation has proposed techniques to generate
\emph{refinement} proofs, typically for compiler optimizations~\cite{pnueli98,
necula2000translation, barrett2005tvoc}.
A translated program, the \emph{target}, refines its \emph{source} if every
behavior of the target is also a behavior of the source.
To close the gap in ensuring the correctness of C-to-Rust translation, adopting
the approach of proving refinement is a promising direction.



In this work, we present an approach to C-to-Rust translation validation based
on foundational refinement proofs.
Specifically, we tackle two challenges:
(1) defining refinement, as no formal semantics defines the behaviors of C
and Rust programs in the same terms;
(2) proving refinement, as C and Rust represent the same data
differently, and the proof must relate the two representations.



\paragraph{\textbf{Challenge \#1.}}

A foundational refinement proof requires formal semantics
that define the behaviors of the source and target programs.
Refinement is an inclusion between the sets of behaviors,
which is well-defined only if both formal semantics define
program behaviors in the same terms,
such as traces of the same kind of events.
However, existing formal semantics for C~\cite{cerberus16, compcert09}
and Rust~\cite{rustbelt18, refinedrust24, verus23}
do not define behaviors in the same terms.
The formal semantics for C and Rust
were developed independently and differ in how they model
values, operations, and memory.




We address this challenge by validating the translation at the
LLVM intermediate representation (LLVM IR) level.
LLVM IR is an intermediate language used by compilers for
many programming languages, including C and Rust.
Its formal semantics has been developed
to verify compiler optimizations through refinement~\cite{vellvm21, alive221}.
We compile the source C program and its Rust translation into LLVM IR
and prove that the compiled Rust program refines the compiled C program.





\paragraph{\textbf{Challenge \#2.}}

C and Rust often represent the same data differently, which makes refinement
proofs challenging.
For example, a variable-length array (VLA) in C is translated into a \C{Vec} in
Rust, which is a struct containing a pointer to a heap-allocated array;
they differ in memory layout and supported operations.
These differences remain even after compilation to LLVM IR, where a C VLA is
represented as a pointer to its first element and a Rust \C{Vec} is represented
as a struct.


Therefore, a refinement proof should establish a relation between data in two
different representations.
Following the usual \emph{simulation} approach, a proof starts the two programs
with related input states, follows their executions, and compares their output
states, where each step involves re-establishing the relation between the data.
This is challenging, as the data is typically modified through library
functions, whose calling conventions and internal logic differ between C and
Rust.



We address this challenge by introducing a \emph{common abstraction} for each
pair of C and Rust representations that frequently correspond in translation,
relating two representations of the same data.
A common abstraction specifies the type of an abstract value shared by the two
representations.
For example, a C VLA and a Rust \C{Vec} have a common abstraction: a list.

We formalize the concept of a common abstraction by defining a predicate over
\emph{ghost resources}~\cite{iris15, iris18} that expresses how a value is laid
out in memory in each representation, so that a proof can reason about memory
updates at the level of abstract values.
With a common abstraction, simulation starts with the same abstract value in two
representations and, at the end, compares the outputs as abstract values.
We also describe the behavior of each library function with a specification over
abstract values, reducing the effort required to re-establish the relation after
each library call.



\paragraph{\textbf{AI-assisted proof automation.}}

Existing translation validation techniques~\cite{pnueli98,
necula2000translation, barrett2005tvoc} construct refinement proofs using
deterministic procedures, often by exploiting known structural correspondences
or transformation patterns in their domains.
However, our translation validation also targets translations produced by
LLM-based translators, which may restructure the code in arbitrary ways to
produce idiomatic Rust code.
For this reason, we instead leverage AI agents to automate proof construction,
taking advantage of recent advances in AI's ability to generate formal proofs~\cite{autoverus25, anthropic2026formal}.
Since a proof assistant checks the generated proofs, the agents need not be trusted.

For efficient proof generation, we construct a refinement proof as a chain of
multiple refinement relations, where only one relation is program-specific while
the others are common to all translations.
The common parts are proved in advance, and only the program-specific part is
proved by an agent at validation time through simulation.
Furthermore, we provide simulation rules that capture common reasoning patterns
required in proofs.


\paragraph{\textbf{Implementation and evaluation.}}

We implement our approach as a tool named \toolname{} based on the Rocq proof
assistant.
Our semantics of LLVM IR is a variant of Vellvm~\cite{vellvm21},
a formal semantics of LLVM IR mechanized in Rocq.
The refinement relations and theorems for proving them
are taken from CRIS~\cite{cris26}, a Rocq framework for refinement proofs.
On top of these, we build our program models,
common abstractions, and specifications of library functions.


We evaluate \toolname{} on LLM translations of the 519 C programs
in the benchmark suite of VERT~\cite{vert25}.
\toolname{} proves the translations of 486 programs correct
without bounding their executions,
and discovers three incorrect translations, providing valid counterexamples
that neither VERT nor RustAssure finds.
It validates a translation in 23.3 minutes at a cost of \$2.91 on average using
\C{gpt-6.1-sol},
and more capable LLMs prove complex programs faster at a higher cost.
We also conduct a case study on QOI~\cite{qoi22}, a small real-world image compression library,
demonstrating how \toolname{} handles complex features such as file operations.

\paragraph{\textbf{Contributions.}}
To summarize, the contributions of this paper are as follows:
\begin{itemize}[leftmargin=1cm]
  \item We formulate C-to-Rust translation validation as a refinement proof
    between the compiled programs at the LLVM IR level (\cref{sec:validation}).
  \item We introduce common abstractions to relate different data
    representations of C and Rust through abstract values and specify the
    behavior of library functions over those values (\cref{sec:common}).
  \item We construct a refinement proof as a chain of refinement relations and
    show how the program-specific part can be proved by simulation
    (\cref{sec:proof}).
  \item We implement the proposed approach in Rocq on top of Vellvm and CRIS,
    automating the construction of refinement proofs with AI agents
    (\cref{sec:automation}).
  \item We evaluate \toolname{} on 519 C programs, proving 486 translations
    correct and finding three incorrect translations, and also conduct a case
    study on an image compression library (\cref{sec:evaluation}).
\end{itemize}

%% file: sections/2-1-llvm.tex
\section{Background}


This section introduces LLVM IR and its semantics (\cref{sec:llvm}) and explains
how CRIS defines refinement (\cref{sec:refinement}).
This and subsequent sections use \C{has\_duplicate} in \cref{fig:example} as a
running example, so we first explain the example here.
The top left shows a C function that takes a pointer \C{a} to an integer array
of length \C{n}, copies the data from \C{a} to a VLA \C{s}, sorts \C{s}, and
checks for a duplicate within a loop.
Note that \C{lib.h} provides \C{void sort(int *first, int *last)}, which sorts
the elements from \C{first} to \C{last}.
The top right shows its Rust translation, which takes a slice \C{a}, represented
as a fat pointer accompanied by its length, copies the data from \C{a} to a
\C{Vec} \C{s}, sorts \C{s}, and checks for a duplicate using methods provided by
Rust iterators.

\subsection{LLVM IR}
\label{sec:llvm}

LLVM IR is a typed language in static single assignment (SSA) form, used by
Clang and rustc.
Each SSA register, whose name starts with \C{\%}, is assigned exactly once.
A program consists of functions and global variables, whose names start with
\C{@}.
The body of a function consists of basic blocks, each of which is a sequence of
instructions ending with a terminator.


\begin{figure}[t]
\begin{minipage}[t]{0.485\textwidth}
\begin{lstlisting}[style=c, emph={has_duplicate, memcpy, sort}]
#include <string.h>
#include "lib.h"
bool has_duplicate(const int *a, size_t n) {
  int s[n];
  memcpy(s, a, sizeof s);
  sort(s, s + n);
  for (size_t i = 1; i < n; i++)
    if (s[i] == s[i - 1])
      return true;
  return false;
}
\end{lstlisting}
\smallskip\hrule\smallskip
\begin{lstlisting}[style=llvm, emph={@has_duplicate,@llvm.memcpy,@sort}]
define i1 @has_duplicate(ptr %a, i64 %n) {
entry:
  %vla = alloca i32, i64 %n
  %0 = shl i64 %n, 2
  call void @llvm.memcpy(ptr %vla, ptr %a, i64 %0,
                         i1 false)
  %add.ptr = getelementptr i32, ptr %vla, i64 %n
  call void @sort(ptr %vla, ptr %add.ptr)
  %cmp = icmp ugt i64 %n, 1
  br i1 %cmp, label %for.body, label %cleanup
for.body:
  %i = phi i64 [1, %entry], [%inc, %for.body]
  %idx = getelementptr i32, ptr %vla, i64 %i
  %1 = load i32, ptr %idx
  %idx1 = getelementptr i8, ptr %idx, i64 -4
  %2 = load i32, ptr %idx1
  %cmp2 = icmp eq i32 %1, %2
  %inc = add i64 %i, 1
  %last = icmp eq i64 %inc, %n
  %stop = select i1 %cmp2, i1 true, i1 %last
  br i1 %stop, label %cleanup, label %for.body
cleanup:
  %r = phi i1 [false, %entry], [%cmp2, %for.body]
  ret i1 %r
}
\end{lstlisting}
\end{minipage}\hfill\vrule\hfill
\begin{minipage}[t]{0.485\textwidth}
\begin{lstlisting}[style=rust, emph={has_duplicate, to_vec, sort, windows, any}]
pub fn has_duplicate(a: &[i32]) -> bool {
  let mut s = a.to_vec();
  s.sort();
  s.windows(2).any(|w| w[0] == w[1])
}
\end{lstlisting}
\smallskip\hrule\smallskip
\begin{lstlisting}[style=llvm, emph={@has_duplicate,@to_vec,@sort,@drop_in_place}]
define i1 @has_duplicate(ptr %a.0, i64 %a.1) {
start:
  ; Vec offsets {cap: 0, ptr: 8, len: 16}
  %s = alloca [24 x i8]
  call void @to_vec(ptr %s, ptr %a.0, i64 %a.1)
  %0 = getelementptr i8, ptr %s, i64 8
  %buf = load ptr, ptr %0
  %1 = getelementptr i8, ptr %s, i64 16
  %len = load i64, ptr %1
  call void @sort(ptr %buf, i64 %len)
  %buf1 = load ptr, ptr %0
  %len1 = load i64, ptr %1
  br label %loop
loop:
  %p = phi ptr [%buf1, %start], [%p1, %body]
  %k = phi i64 [%len1, %start], [%k1, %body]
  %more = icmp ugt i64 %k, 1
  br i1 %more, label %body, label %exit
body:
  %k1 = add i64 %k, -1
  %p1 = getelementptr i8, ptr %p, i64 4
  %x = load i32, ptr %p
  %2 = getelementptr i8, ptr %p, i64 4
  %y = load i32, ptr %2
  %eq = icmp eq i32 %x, %y
  br i1 %eq, label %exit, label %loop
exit:
  call void @drop_in_place(ptr %s)
  ret i1 %more
}
\end{lstlisting}
\end{minipage}
\caption{A C function \texttt{has\_duplicate} (left) and its Rust translation (right),
each above the LLVM IR that Clang or rustc generates from it at O1.
We omit attributes, flags, alignment, and metadata, and rename registers.}
\label{fig:example}
\end{figure}

\cref{fig:example} shows LLVM IR programs compiled from \C{has\_duplicate}:
the bottom left is compiled from C, and the bottom right is compiled from Rust.
Both the C VLA \C{s} and the Rust \C{Vec} \C{s} become stack memory allocated by
\C{alloca}:
for the VLA, an array is allocated and pointed to by \C{\%vla};
for the \C{Vec}, a struct that later holds a pointer to a heap buffer is
allocated and pointed to by \C{\%s}.
During compilation, library function calls are handled in three ways:
(1) compiled to an LLVM \emph{intrinsic} call, e.g., \C{memcpy} becomes
\C{@llvm.memcpy};
(2) compiled to an external function call, e.g., \C{sort} and \C{to\_vec};
or (3) inlined, e.g., \C{windows} and \C{any} are inlined, so the basic blocks
form a loop.

%

LLVM IR has two mechanisms for introducing \emph{undefined behavior}
(UB)~\cite{tamingub17}:
some operations, such as division by zero, \emph{immediately} cause UB;
others, such as overflow of an addition marked as non-wrapping, \emph{defer} UB
by returning a special value, namely \C{poison}, which propagates
through most operations and causes actual UB only when its concrete value is
required, e.g., in a branch condition.
Another special value for deferred UB is \C{undef}, which represents nondeterminism.
The LLVM community is working to remove \C{undef} from LLVM IR
entirely~\cite{llvmundef26}, so our semantics only includes \C{poison} and
excludes \C{undef};
since current compilers may still produce IR with \C{undef}, we eliminate
\C{undef} from compiled IR (\cref{sec:automation}).


%

The denotational semantics of LLVM IR is defined using \emph{interaction trees}
(ITrees)~\cite{itree19}, which represent computations that interact with their
environment through events.
ITrees can express executions of any length, including nonterminating ones,
because they are coinductively defined:
\[\small
\begin{array}[t]{@{}l@{\ }l@{}}
\multicolumn{2}{@{}l@{}}{\mathsf{itree}\ E\ R \triangleq} \\
\mid & \mathsf{Ret}\ (r : R) \\
\mid & \mathsf{Tau}\ (T : \mathsf{itree}\ E\ R) \\
\mid & \mathsf{Vis}\ \{X\}\ (\mathit{ev} : E\,X)\ (k : X \to \mathsf{itree}\ E\ R)
\end{array}
\qquad
\begin{array}[t]{@{}l@{}}
x \leftarrow T_1;\ T_2 \triangleq \mathsf{bind}\ T_1\ (\lambda x.\ T_2) \\
\quad\ \mcomment{run $T_1$: itree, then $T_2$: itree with $x$ = return of $T_1$} \\
\mathsf{trigger}\ \mathit{ev} \triangleq \mathsf{Vis}\ \mathit{ev}\ (\lambda x.\ \mathsf{Ret}\ x) \\
\quad\ \mcomment{trigger the event $\mathit{ev}$ and return the response}
\end{array}
\]
An ITree either returns a value ($\mathsf{Ret}$), takes a silent step
($\mathsf{Tau}$), or triggers an event and continues with the response
($\mathsf{Vis}$).
For events, we follow the definition in CRIS:
{\small\[
\begin{array}{@{}r@{\;}c@{\;}l@{}}
E & ::= & \mathsf{Call}(f, \mathit{v}) \mid \mathsf{IO}(f, \mathit{v}) \mid \mathsf{Choose}\ X \mid \mathsf{UB} \mid \mathsf{NB} \\
  & \mid & \imop{Take}\ X \mid \imop{Assume}\ Q \mid \imop{Guarantee}\ Q \quad \mcomment{imaginary operations }
\end{array}
\]
An event is either a function call ($\mathsf{Call}$), an interaction with the
environment ($\mathsf{IO}$) with a name $f$ and an argument $\mathit{v}$,
a nondeterministic choice ($\mathsf{Choose}$), UB
($\mathsf{UB}$), no behavior ($\mathsf{NB}$), or an \emph{imaginary operation} of CRIS
($\imop{Take}$, $\imop{Assume}$, or $\imop{Guarantee}$,
where $Q$ is an assertion on ghost resources).
Imaginary operations are not used in the LLVM IR semantics but are used in our
specifications of library functions (\cref{sec:chain}).



Our semantics, given in detail in \cref{app:semantics}, is a variant of
Vellvm~\cite{vellvm21}, with two changes: it uses CRIS's events, and it
passes memory directly instead of modeling memory operations as events.
The denotation of an instruction $i$, $\sem{i}(\rho, m) : \mathsf{itree}\ E\
(\mathit{Val} \times \mathit{Mem})$, takes a local environment $\rho$, which
maps registers to values, and a memory $m$, which is a collection of allocated
blocks of bytes, and returns an ITree that produces a value and a new memory.
Similarly, $\sem{e}(\rho, m) : \mathsf{itree}\ E\ \mathit{Val}$
evaluates an expression $e$ to a value,
and $\sem{\mathit{es}}$ evaluates a list $\mathit{es}$ of expressions from left to right.
For example, the semantics of $\mathsf{load}$ and $\mathsf{call}$ instructions
is as follows:
\[\small
\begin{array}{@{}r@{\;}c@{\;}l@{}}
\sem{\mathsf{load}(\tau, e)}(\rho, m) & = &
  p \leftarrow \sem{e}(\rho, m);\ \mathbf{match}\ \pf{\mathsf{load}}(\tau, p, m)\ \mathbf{with} \\
& & \phantom{p \leftarrow \sem{e}(\rho, m);\ }\quad
  \mathsf{Success}\ v \Rightarrow \mathsf{Ret}\ (v, m) \mid \mathsf{Failure} \Rightarrow \mathsf{trigger}\ \mathsf{UB} \\[2pt]
\sem{\mathsf{call}(\tau, e, \mathit{es})}(\rho, m) & = &
  \mathit{args} \leftarrow \sem{\mathit{es}}(\rho, m);\ p \leftarrow \sem{e}(\rho, m);\
  \mathbf{if}\ p = \texttt{poison}\ \mathbf{then}\ \mathsf{trigger}\ \mathsf{UB} \\
& & \phantom{\mathit{args} \leftarrow \sem{\mathit{es}}(\rho, m);\ p \leftarrow \sem{e}(\rho, m);\ }
  \mathbf{else}\ \mathsf{trigger}\ \mathsf{Call}(\mathit{fn}(p), (\tau, \mathit{args}, m))
\end{array}
\]
$\mathsf{load}(\tau, e)$ evaluates an expression $e$ to obtain a pointer and
reads a value of type $\tau$ from the pointed-to memory location
using a pure function $\pf{\mathsf{load}}$, shown in \pf{blue},
which does not trigger any events.
$\pf{\mathsf{load}}$ returns $\mathsf{Success}\ v$ with the read value $v$,
or $\mathsf{Failure}$ for an invalid access, on which the denotation triggers $\mathsf{UB}$.
$\mathsf{call}(\tau, e, \mathit{es})$ evaluates $\mathit{es}$ and $e$ to obtain
arguments and a function pointer, and triggers a $\mathsf{Call}$ event
(or $\mathsf{UB}$ if the pointer is \C{poison}), passing
the function pointer, the arguments, and the current memory, and receiving the result of type $\tau$ and the new
memory.
The denotation of an LLVM IR function $F$, $\sem{F}(\mathit{args}, m) :
\mathsf{itree}\ E\ (\mathit{Val} \times \mathit{Mem})$, takes the arguments and
the initial memory and returns an ITree that produces the return value and the
final memory.
It executes the basic blocks starting from the entry block.



%% file: sections/2-2-refinement.tex
\subsection{Refinement in CRIS}
\label{sec:refinement}


In CRIS, a \emph{module} is a mapping from function names to function bodies,
where a function body takes a Rocq value and returns an ITree producing a Rocq
value:
$\small \mathit{Mod} \; \triangleq \; \mathit{Fname} \rightharpoonup
(\mathit{Any} \to \mathsf{itree}\ E\ \mathit{Any})$.

A $\mathsf{Call}$ event to a function within the same module is handled by the
function's body, while calling a function not in the module causes UB.
To call a function in another module, modules $M_1$ and $M_2$ can be linked to
form a module $M_1 \star M_2$, which contains functions from both modules:
$\small M_1 \star M_2 \; \triangleq \; M_1 \cup M_2$.


A \emph{program} is a module that designates one of its functions that does not
use its input as the \emph{entry}.
The denotation of a program $M$ is $\sem{M} : \mathsf{itree}\ E'\ \mathit{Any}$, obtained from the ITree
returned by the entry function by replacing every $\mathsf{Call}$ event with the
body of the callee,
where $E'$ consists of events other than $\mathsf{Call}$.

%
%
%
%

\begin{figure}[t]
\small
\[
\begin{array}{@{}l@{}}
\mathit{Trace} \triangleq
  \{\mathsf{Interact}\ (f, a, r) :: \mathit{tr} \mid f \in \mathit{string},\ a, r \in \mathit{Any},\ \mathit{tr} \in \mathit{Trace}\}
  \uplus \{\mathsf{Done}\ v \mid v \in \mathit{Any}\} \\[4pt]
\mathsf{Beh}(T \in \mathsf{itree}\ E'\ \mathit{Any}) \in \mathcal{P}(\mathit{Trace}) \triangleq
  \mathbf{match}\ T\ \mathbf{with} \quad \mcomment{behavior}\\
\quad
\begin{array}{@{}l@{\qquad}l@{}}
\mid \mathsf{Ret}\ v \Rightarrow \{\mathsf{Done}\ v\}
  & \mid \mathsf{Tau}\ T' \Rightarrow \mathsf{Beh}(T') \\
\mid \mathsf{Vis}\ \mathsf{IO}(f, a)\ k \Rightarrow
  \bigcup_{r} \mathsf{Interact}\ (f, a, r) :: \mathsf{Beh}(k\ r)
  & \mid \mathsf{Vis}\ \mathsf{UB}\ k \Rightarrow \mathit{Trace} \\
\mid \mathsf{Vis}\ (\mathsf{Choose}\ X)\ k \Rightarrow \bigcup_{x \in X} \mathsf{Beh}(k\ x)
  & \mid \mathsf{Vis}\ \mathsf{NB}\ k \Rightarrow \emptyset \\
\mid \mathsf{Vis}\ (\imop{Assume}\ Q)\ k \Rightarrow
    \mathbf{if}\ Q\ \mathbf{then}\ \mathsf{Beh}(k\ ())\ \mathbf{else}\ \mathit{Trace}
  & \mid \mathsf{Vis}\ (\imop{Take}\ X)\ k \Rightarrow \bigcap_{x \in X} \mathsf{Beh}(k\ x) \\
\mid \mathsf{Vis}\ (\imop{Guarantee}\ Q)\ k \Rightarrow
    \mathbf{if}\ Q\ \mathbf{then}\ \mathsf{Beh}(k\ ())\ \mathbf{else}\ \emptyset
\end{array} \\
\mathbf{end} \\[4pt]
\begin{array}{@{}l@{\;}c@{\;}l@{\quad}l@{}}
M \sqsubseteq_{\mathrm{beh}} M' & \triangleq &
  \mathsf{Beh}(\sem{M}) \subseteq \mathsf{Beh}(\sem{M'}) &
  \mcomment{behavioral refinement}
\end{array}
\end{array}
\]
\caption{Definitions of trace, behavior, and refinement.}
\label{fig:refinement}
\end{figure}

\cref{fig:refinement} defines the behavior of a program $M$ as a set of
\emph{traces}, $\mathsf{Beh}(\sem{M})$.
A trace records the IO interactions of an execution ($\mathsf{Interact}$), each with its name $f$,
argument $a$, and response $r$, and ends with $\mathsf{Done}\ v$, where $v$ is the value
returned by the entry function.
$\mathsf{Beh}(T)$ collects the traces of a tree $T$ by cases:
a returned value $v$ gives a trace $\mathsf{Done}\ v$,
a silent step $\mathsf{Tau}$ is skipped,
$\mathsf{IO}$ gives traces for all responses,
$\mathsf{Choose}$ gives traces for all choices,
$\mathsf{UB}$ gives every trace, and $\mathsf{NB}$ gives none.
For the imaginary operations,
$\imop{Take}$ gives the traces common to all choices,
and $\imop{Assume}\ Q$ and $\imop{Guarantee}\ Q$ continue if $Q$ holds
and otherwise behave as $\mathsf{UB}$ and $\mathsf{NB}$, respectively.
A program $M$ refines $M'$, written $M \sqsubseteq_{\mathrm{beh}} M'$, if every
trace of $\sem{M}$ is a trace of $\sem{M'}$.%
\footnote{%
  The actual definition of $\sqsubseteq_{\mathrm{beh}}$ in CRIS also runs the two programs with ghost resources:
  $M$ with the empty resource and $M'$ with some valid
  initial resource. Resources affect behaviors only through the imaginary
  operations, so for programs without them, the two definitions coincide.
  We also omit traces not ending with $\mathsf{Done}$, such as divergence and abort.}

%% file: sections/3-validation.tex
\section{Defining Translation Correctness as Refinement}
\label{sec:validation}

In this section, we define the correctness of C-to-Rust translation as
refinement in CRIS at the LLVM IR level.
We represent a C program and its translated Rust program as CRIS programs and
define translation correctness as refinement between these two CRIS programs.
We restrict our scope to programs with a single public function;
in general, when a program has multiple public functions, translation validation
could be performed for each pair of corresponding public functions.
We first construct a CRIS function from an LLVM IR function
(\cref{sec:programs}), then build a CRIS program from the CRIS function by
linking it with library modules (\cref{sec:library}), and finally define
correctness as refinement between CRIS programs (\cref{sec:correctness}).


\subsection{Constructing CRIS Functions}
\label{sec:programs}

We construct a CRIS function, a \emph{driver}, for each LLVM IR function.
When building a CRIS program later, each driver serves as the entry of the
program.
The goal of drivers is to resolve the discrepancies between C and Rust, enabling
refinement proofs.
Refinement is defined as a subset relation between trace sets, and traces carry
values.
Therefore, input and output values should be represented identically for both
LLVM IR functions compiled from C and Rust.
Drivers achieve this goal by taking abstract values as input, converting the
input to the representation of each language, running the LLVM IR function with
the correct representation, and converting the results back to abstract values.
For an LLVM IR function $F_X$ from each language $X \in \{\mathit{C},
\mathit{Rust}\}$, its driver $\mathit{driver}(F_X)$ is as follows:
\[\footnotesize
\begin{array}{@{}r@{\quad}l@{\quad}|@{\quad}r@{\ }l@{}}
  \text{\scriptsize 1} & \mathit{driver}(F_X) \triangleq \lambda\_.
    & & \\
  \text{\scriptsize 2} & \quad v \leftarrow \mathsf{trigger}\ \mathsf{IO}(\texttt{"input"}, ());
    & \pf{valid\_input}_F : & \mathit{list}\ \mathit{Any} \to \mathit{Prop} \\
  \text{\scriptsize 3} & \quad \mathbf{if}\ \neg\,\pf{valid\_input}_F(v)\ \mathbf{then}\ \mathsf{Ret}\ \mathsf{None}
    & \pf{register}_{F,X} : & \mathit{list}\ \mathit{Any} \times \mathit{Mem} \to \mathit{list}\ \mathit{Obj} \times \mathit{Mem} \\
  \text{\scriptsize 4} & \quad \mathbf{else}\ (p, m) \leftarrow \pf{register}_{F,X}(v, m_0);
    & \pf{args}_{F,X} : & \mathit{list}\ \mathit{Any} \times \mathit{list}\ \mathit{Obj} \to \mathit{list}\ \mathit{Val} \\
  \text{\scriptsize 5} & \quad \phantom{\mathbf{else}\ } (r, m') \leftarrow \sem{F_X}(\pf{args}_{F,X}(v, p), m);
    & \pf{read}_{F,X} : & \mathit{list}\ \mathit{Obj} \times \mathit{Mem} \to \mathit{list}\ \mathit{Any} \\
  \text{\scriptsize 6} & \quad \phantom{\mathbf{else}\ } o \leftarrow \mathit{observe}(r);\ o' \leftarrow
  \mathit{observe}(\pf{read}_{F,X}(p, m'));
    & \mathit{observe} : & A \to \mathsf{itree}\ E\ A \\
  \text{\scriptsize 7} & \quad \phantom{\mathbf{else}\ } \mathsf{Ret}\ (\mathsf{Some}\ (o, o'))
    & &
\end{array}
\]

\vspace{-\parskip}
As the driver is the entry of a CRIS program, it does not use its argument
(line~1) and instead receives the input $v$ by triggering an $\mathsf{IO}$ event
(line~2).
By taking the input through an event, the driver records it in a trace.
The input is a list of abstract values, whose types are determined by the common
abstractions of the arguments of the C and Rust functions.
For example, the drivers for \C{has\_duplicate} take a single abstract value,
a list of 32-bit signed integers.

Then, the driver determines whether $v$ is an intended input to the function $F$
using $\pf{valid\_input}_F$;
if not, it immediately returns $\mathsf{None}$ (line~3).
We allow the user to specify the intended input space (\cref{sec:automation}).
For \C{has\_duplicate}, it checks whether the length of the list is within the
size bound of the architecture.

If the input is valid, the driver converts the input to the appropriate
representation for the language (line~4) and runs the LLVM IR function using its
denotation $\sem{F_X}$ (line~5):
$\pf{register}_{F,X}$ allocates $v$ in the initial memory $m_0$ using the
representation of language $X$, returning a list
of objects, each of which is an LLVM IR value or a tuple of such values,
that hold information about the registered input, and the updated memory $m$;
$\pf{args}_{F,X}$ constructs the arguments required by the LLVM IR function.
For $\C{has\_duplicate}$, the list is allocated as an array, and $p$ contains the
pointer to the array and its length;
the arguments are these two values, passed as parameters
\C{\%a} and \C{\%n} for C and as \C{\%a.0} and \C{\%a.1} for Rust.

After the call, the driver collects the output (line~6).
It takes the return value $r$ of the function and, to also capture observable
memory effects, reads the data referred to by the pointer arguments as
abstract values using $\pf{read}_{F,X}$.
For \C{has\_duplicate}, it returns the list represented by the array pointed to
by $p$.

The driver \emph{observes} each output value to properly handle \C{poison}:
when an LLVM IR function compiled from C produces \C{poison}, a function
compiled from Rust should be considered to refine it regardless of the value it
produces.
For this purpose, we introduce \emph{value refinement}:
$v$ refines $v'$, written $v \sqsubseteq v'$, if $v = v'$, or if $v'$ is
\C{poison} and $v$ has the same type;
aggregate values, including abstract values such as lists, are compared element-wise.
$\mathit{observe}(v)$ provides any non-\C{poison} value that refines $v$:
\[\small
\mathit{observe}(v) \triangleq
\mathsf{trigger}\ \mathsf{Choose}\ \{x \mid x \sqsubseteq v \land x \neq \C{poison}\}
\]
The driver finishes by returning the observed values, recording the output in a
trace (line~7).

\subsection{Building CRIS Programs}
\label{sec:library}

To build a CRIS program, we need library modules that model the behavior of
library functions, as the translated program may call library functions.
We introduce three modules: $\mathit{lib_{LLVM}}$, $\mathit{lib_{C}}$, and
$\Wfun(\mathit{lib_{Rust}})$, which consist of CRIS functions that express the
behavior of LLVM intrinsics, C library functions, and Rust library functions,
respectively.
For example, \C{sort} in $\mathit{lib_{C}}$ takes two pointers along with the
current memory and returns an ITree that produces the new memory in which the
array between the two pointers is sorted.
Note that $\Wfun(\mathit{lib_{Rust}})$ is a wrapper module generated by CRIS
from the underlying module $\mathit{lib_{Rust}}$ by inserting pre- and
postconditions into ITrees, as explained further in \cref{sec:chain}.
Currently, these modules only include functions used by the benchmark suite in
\cref{sec:evaluation}, but we can model more functions to support additional
programs.

With these library modules, we build a CRIS program from a driver.
We first introduce a main module $\mathit{main}(F_X)$ that consists only of the
driver function $\mathit{driver}(F_X)$.
We then link the main module with the necessary library
modules---$\mathit{lib_{LLVM}}$ and the library module for the relevant
language $L$---to obtain a CRIS program whose entry is the driver in the main
module:
\[\small \mathit{program}(F_X,L) \; \triangleq \; \mathit{main}(F_X) \star L
\star \mathit{lib_{LLVM}} \]
The programs from C and Rust are $\mathit{program}(F_C,\mathit{lib_{C}})$ and
$\mathit{program}(F_{Rust}, \Wfun(\mathit{lib_{Rust}}))$, respectively.

\subsection{Translation Correctness}
\label{sec:correctness}

We define the correctness of translation as follows:

\begin{definition}[Translation Correctness]
\label{def:correctness}
  $F_{\mathit{Rust}}$ is a correct translation of $F_{\mathit{C}}$ if
\[
\mathit{program}(F_{Rust}, \Wfun(\mathit{lib_{Rust}})) \behref \mathit{program}(F_C,\mathit{lib_{C}}).
\]
\end{definition}

\noindent
Our validation aims to prove this refinement, implying that the translation is
correct with respect to our library modules, and thus relies on the assumptions
that these models faithfully represent the actual libraries.
We formally express these assumptions using the notion of \emph{contextual
refinement}:
a module $M$ contextually refines a module $M'$, written $M \ctxref M'$, if
replacing $M'$ with $M$ in any program adds no behavior~\cite{ccr23}:
\[\small M \ctxref M' \;\triangleq\; \forall C.\ M \star C\behref M'\star C\]
Let $\mathit{lib_{LLVM,std}}$, $\mathit{lib_{C,std}}$, and
$\mathit{lib_{Rust,std}}$ be \emph{conceptual} standard library modules that precisely
describe all possible behaviors of the respective libraries.
We assume the following:
\[
  \mathit{lib_{LLVM}}\ctxref\mathit{lib_{LLVM,std}}\qquad
  \mathit{lib_{LLVM,std}}\ctxref\mathit{lib_{LLVM}}\qquad
  \mathit{lib_{C}}\ctxref\mathit{lib_{C,std}}\qquad
  \mathit{lib_{Rust,std}}\ctxref\mathit{\Wfun(lib_{Rust})}
\]
Under these assumptions, \cref{def:correctness} implies correctness with respect
to the actual libraries:
\[
  \begin{array}{@{}c@{\;}c@{\;}c@{\;}c@{\;}c@{\;}c@{\;}c@{\;}l@{}}
    & \mathit{main}(F_{Rust}) &\star& \mathit{lib_{Rust,std}} &\star& \mathit{lib_{LLVM,std}} \\
    \behref & \mathit{main}(F_{Rust}) &\star& \Wfun(\mathit{lib_{Rust}}) &\star& \mathit{lib_{LLVM}}
    & = & \mathit{program}(F_{Rust}, \Wfun(\mathit{lib_{Rust}})) \\
    \behref & \mathit{main}(F_C) &\star& \mathit{lib_{C}} &\star& \mathit{lib_{LLVM}}
    & = & \mathit{program}(F_C,\mathit{lib_{C}}) \\
    \behref & \mathit{main}(F_C) &\star& \mathit{lib_{C,std}} &\star& \mathit{lib_{LLVM,std}}
  \end{array}
\]

%

\vspace{-\parskip}
Furthermore, while our validation is performed at the LLVM IR level, the
refinement extends to the binary and source levels if we assume that the
compilers are correct.
If an LLVM compiler is correct, a compiled binary refines its source LLVM IR;
if a C compiler is correct, the compiled LLVM IR refines its C source.
With such compilers, the binary compiled from the translated Rust code refines
the C source:
\[ \text{Binary from Rust} \behref \text{LLVM IR from Rust} \behref \text{LLVM
IR from C} \behref \text{C source} \]

\vspace{-\parskip}
However, validation at the LLVM IR level is incomplete:
even when the translated Rust code refines the C source, the LLVM IR from Rust
may not refine the LLVM IR from C, as the LLVM IR does not necessarily preserve
all possible behaviors of the source.
For example, a compiler may turn UB in the C source into defined behavior in the
LLVM IR, which is legitimate since the LLVM IR only needs to refine the source.
We discuss cases where validation fails for this reason in
\cref{sec:rq1}.

%% file: sections/4-common.tex
\section{Reasoning with Common Abstractions}
\label{sec:common}

We now introduce common abstractions, which we use to reason about different
representations of the same abstract value in C and Rust in refinement proofs.
Proving refinement requires establishing a simulation between the two programs,
and we represent common abstractions with ghost resources~\cite{iris15, iris18}
that relate the memories of the two programs.
During simulation, resources expressing a common abstraction are obtained when
the input is allocated, and they are maintained throughout execution, even
across library function calls, until the results are read back.
We introduce \emph{representation predicates}, which describe an abstract value
in a specific representation (\cref{sec:ownership}), and common abstractions,
which connect representation predicates of C and Rust (\cref{sec:ca}), and
provide specifications of library functions in terms of common abstractions
(\cref{sec:spec}).

%


\subsection{Representation Predicates}
\label{sec:ownership}

We reason about memory using ghost resources, which do not affect
execution, with ownership of cells as the most basic resource.
We write $p \mapsto_{\dq} u$ for the ownership of the cells at $p$
that hold an LLVM IR value $u$, with right $\dq$; the full right $\dq = 1$ allows writing,
and a smaller right allows only reading.
The separating conjunction~\cite{seplogic01} $P \ast Q$ holds when the resources
can be split into two disjoint parts,
one satisfying $P$ and the other satisfying $Q$.

To describe the actual memory,
we also introduce a \emph{memory authority} $\mathsf{Auth}(m)$,
which tracks the entire memory $m$.
Every ownership resource must agree with it:
$\mathsf{Auth}(m) \ast p \mapsto_{\dq} u$ ensures that
$p$ points to live cells in $m$ that hold $u$.
When the memory changes, $\mathsf{Auth}(m)$ is updated to reflect the
new memory.
The C and Rust programs run on separate memories, each with its own authority.


Since C and Rust may lay out the same abstract value differently,
we describe each layout with a representation predicate
defined using ownership resources.
A representation predicate $\rep[\dq]{a}{R}{v}$ is a resource
stating that an LLVM IR value $a$
represents an abstract value $v$ in a representation $R$, with right $\dq$;
$a$ can be a tuple of values, such as a pointer and a length.
We define the predicate for each kind of abstract value
and each representation.
For example, in \C{has\_duplicate}, a list stored in the local variable \C{s} has
two different representations, $\mathsf{Array}$ in C and $\mathsf{Vec}$ in Rust,
whose predicates are defined as follows:
\[ \small
\begin{array}{@{}r@{\;}c@{\;}l@{}}
\rep[\dq]{(p, n)}{(\mathsf{Array}\ \tau)}{\mathit{vs}} & \triangleq &
  \pure{n = |\mathit{vs}|} \ast
  \bigsep_{0 \le i < n}\, (p + i \cdot |\tau|) \mapsto_{\dq} \mathit{vs}[i] \\
\rep[\dq]{h}{(\mathsf{Vec}\ \tau)}{\mathit{vs}} & \triangleq &
  \exists\, c\, d.\ h \mapsto_{\dq} (c, d, |\mathit{vs}|) \ast
  \rep[\dq]{(d, |\mathit{vs}|)}{(\mathsf{Array}\ \tau)}{\mathit{vs}}
\end{array}
\]
For $\mathsf{Array}$, $\pure{n = |\mathit{vs}|}$ states the pure fact that
$n$ equals the length of the list,
and the $i$-th element is stored at offset $i \cdot |\tau|$ from $p$,
where $|\tau|$ is the size of the element type $\tau$.
For $\mathsf{Vec}$, the header struct $h$ holds the capacity $c$, the address $d$
of the buffer, and the length $|\mathit{vs}|$,
and $\rep[\dq]{(d, |\mathit{vs}|)}{(\mathsf{Array}\ \tau)}{\mathit{vs}}$
states that the first $|\mathit{vs}|$ cells of the buffer hold $\mathit{vs}$.
Thus, a list $\mathit{vs}$ in \C{has\_duplicate} is described by
$\rep[1]{(\reg{vla}, \reg{n})}{(\Arr{i32})}{\mathit{vs}}$
in C and by
$\rep[1]{\reg{s}}{(\Vect{i32})}{\mathit{vs}}$
in Rust.

%

\subsection{Common Abstractions}
\label{sec:ca}


When a C object and a Rust object represent the same abstract
value up to refinement, we say that they have a common abstraction.
For example, a C VLA and a Rust \C{Vec},
described by $\mathsf{Array}$ and $\mathsf{Vec}$,
have the common abstraction List when they represent the same list.
\cref{tab:common-abstractions} enumerates the common abstractions we support.
We identified them from the benchmark suite in \cref{sec:evaluation}, and this
list can be extended to support more abstractions.

\begin{table}[t]
\small
\caption{Supported common abstractions.}
\label{tab:common-abstractions}
\begin{tabular}{@{}lll@{}}
\toprule
Common abstraction (abstract value) & C representation & Rust representation \\
\midrule
List (list of elements) & \texttt{T*} with a length, \texttt{T[n]}, VLA
  & \texttt{\&[T]}, \texttt{\&mut [T]}, \texttt{[T; n]}, \texttt{Vec<T>} \\
String (list of ASCII bytes) & NUL-terminated \texttt{char*} & \texttt{\&str}, \texttt{Vec<char>} \\
Struct (record of fields) & \texttt{struct} (non-recursive) & \texttt{struct} (non-recursive) \\
Table (list of lists) & two-dimensional VLA & \texttt{Vec<Vec<T>{}>} \\
File (file state) & \texttt{FILE*} & \texttt{File} \\
\bottomrule
\end{tabular}
\end{table}

A \emph{representation pair}, or a pair for short, is a resource stating that
a C object and a Rust object have a common abstraction.
A pair $\pair[\dq]{a_c}{R_c}{a_r}{R_r}(v_c)$ states that
the C object $a_c$ represents $v_c$ in representation $R_c$,
and the Rust object $a_r$ represents $v_r$ that refines $v_c$
in representation $R_r$:
\[\small
\pair[\dq]{a_c}{R_c}{a_r}{R_r}(v_c) \triangleq
  \exists\, v_r.\ \rep[\dq]{a_c}{R_c}{v_c} \ast \rep[\dq]{a_r}{R_r}{v_r}
  \ast \pure{v_r \sqsubseteq v_c}
\]

\vspace{-\parskip}
During simulation, a representation pair is typically obtained when the input is
allocated.
For a function $F$ whose input is a single abstract value
(we write a singleton list $[x]$ as $x$), its drivers $\mathit{driver}(F_{\mathit{C}})$ and
$\mathit{driver}(F_{\mathit{Rust}})$ register the same input in their respective
memories using $\pf{register}_{F,X}$, and we reason about how registration
updates the memories at the level of abstract values using a representation pair
with the following lemma:

\begin{lemma}[Registration]
\label{lem:registration}
For every abstract value $v$ of a common abstraction and memories $m_c$ and $m_r$,
if $\pf{register}_{F,\mathit{C}}(v, m_c) = (a_c, m_c')$
and $\pf{register}_{F,\mathit{Rust}}(v, m_r) = (a_r, m_r')$, then
\[
\mathsf{Auth}(m_c) \ast \mathsf{Auth}(m_r) \Rrightarrow
\mathsf{Auth}(m_c') \ast \mathsf{Auth}(m_r') \ast \pair[1]{a_c}{R_c}{a_r}{R_r}(v),
\]
  where $\pf{register}_{F,\mathit{C}}$ allocates $v$ in representation $R_c$,
  and $\pf{register}_{F,\mathit{Rust}}$ allocates $v$ in representation $R_r$.
  Here, $P \Rrightarrow Q$ means that $P$ can be updated to $Q$
  by changing only ghost resources.
\end{lemma}

\noindent
For example, in \C{has\_duplicate}, the input list $\mathit{vs}$ is allocated as
an array in both the C and Rust drivers, and the registration lemma provides
$\pair[1]{(\reg{a}, \reg{n})}{\Arr{i32}}{(\reg{a.0},
\reg{a.1})}{\Arr{i32}}(\mathit{vs})$.


%

If the obtained representation pair is maintained until the end, we can reason
about the final memories and conclude that the value read by the Rust driver
using $\pf{read}_{F,X}$ refines that read by the C driver:

\begin{lemma}[Read]
\label{lem:observation}
For every abstract value $v$ of a common abstraction and memories $m_c$ and $m_r$,
\[
\mathsf{Auth}(m_c) \ast \mathsf{Auth}(m_r) \ast \pair[\dq]{a_c}{R_c}{a_r}{R_r}(v)
\Rrightarrow
\pure{\pf{read}_{F,\mathit{Rust}}(a_r, m_r) \sqsubseteq \pf{read}_{F,\mathit{C}}(a_c, m_c)},
\]
  where $\pf{read}_{F,\mathit{C}}$ reads $a_c$ in representation $R_c$,
  and $\pf{read}_{F,\mathit{Rust}}$ reads $a_r$ in representation $R_r$.
\end{lemma}

\noindent
For an input of multiple abstract values, the two lemmas apply to each abstract
value allocated in memory, yielding a representation pair for each.

\subsection{Library Specifications}
\label{sec:spec}

We provide the specification of each library function, consisting of its pre- and
postconditions, at the level of abstract values using representation predicates,
to express that corresponding C and Rust functions perform the same operation.
Although the two functions take different arguments
and may represent their objects differently,
their specifications describe the same update to the same abstract value.
%
%
A specification is a Hoare triple $\forall x.\ \{P\}\ f(a)\ \{Q\}$
with precondition $P$ and postcondition $Q$:
for every $x$, a call to $f$ with arguments $a$ requires the resources $P$
and, when it returns, provides the resources $Q$.
%

\begin{figure}[t]
\footnotesize
\[
\begin{array}{@{}l@{}}
\mcomment{C: \texttt{void sort(int *first, int *last)} in \texttt{lib.h}} \\
\forall \mathit{vs}.\
  \{\pure{\reg{last} = \reg{first} + 4 \cdot |\mathit{vs}|}
    \ast \rep[1]{(\reg{first}, |\mathit{vs}|)}{(\Arr{i32})}{\mathit{vs}}\} \\
\qquad \mathtt{@sort}(\reg{first}, \reg{last}) \\
\qquad \{\exists\, \mathit{ws}.\ \rep[1]{(\reg{first}, |\mathit{vs}|)}{(\Arr{i32})}{\mathit{ws}}
    \ast \pure{\mathit{sorted}\ \mathit{ws} \land \mathit{perm}\ \mathit{vs}\ \mathit{ws}}\} \\[6pt]
\mcomment{Rust: \texttt{<[i32]>::sort(\&mut self)}} \\
\forall \mathit{vs}.\
  \{\rep[1]{(\reg{ptr}, \reg{len})}{(\Arr{i32})}{\mathit{vs}}\} \\
\qquad \mathtt{@sort}(\reg{ptr}, \reg{len}) \\
\qquad \{\exists\, \mathit{ws}.\ \rep[1]{(\reg{ptr}, \reg{len})}{(\Arr{i32})}{\mathit{ws}}
    \ast \pure{\mathit{sorted}\ \mathit{ws} \land \mathit{perm}\ \mathit{vs}\ \mathit{ws}}\}
\end{array}
\]
\caption{The specifications of \texttt{sort} in C and Rust.}
\label{fig:sort-specs}
\end{figure}

The specifications of \C{sort} in \cref{fig:sort-specs}
describe the same operation on the same abstract value,
although the C and Rust functions take different arguments.
The C function takes pointers to the first element and one past the last element,
whereas the Rust function takes a pointer and a length;
both preconditions describe the array specified by these arguments
as a list $\mathit{vs}$ in the representation $\Arr{i32}$.
Both postconditions state that the same array then represents
a sorted permutation $\mathit{ws}$ of $\mathit{vs}$.

For C library functions, we prove that the specifications are sound with respect
to their bodies, so these specifications do not need to be trusted.
In contrast, for Rust library functions, our assumption
$\mathit{lib_{Rust,std}}\ctxref\Wfun(lib_{Rust})$ states that the specifications
are trusted, as they are integrated into $\Wfun(lib_{Rust})$ and determine its
behavior.
We do not provide specifications for LLVM intrinsics.

To facilitate reasoning about common abstractions across library calls,
we introduce Rule \textsc{Pair-Update}, which captures that
when two corresponding calls that return without triggering $\mathsf{IO}$ events update the objects of a representation pair in the same way,
the pair still holds after the calls, with the updated value:
\[\footnotesize
\inferrule*[right=\textsc{Pair-Update}]
  {\forall v.\ \{\rep[1]{a_c}{R_c}{v}\}\ \mathit{call}_c\ \{\rep[1]{a_c}{R_c}{(f_c\, v)}\}
   \quad \forall v.\ \{\rep[1]{a_r}{R_r}{v}\}\ \mathit{call}_r\ \{\rep[1]{a_r}{R_r}{(f_r\, v)}\} \\\\
   \mathit{call}_c, \mathit{call}_r \text{ return without triggering } \mathsf{IO} \text{ events}
   \quad \forall v_c\, v_r.\ v_r \sqsubseteq v_c \implies f_r\, v_r \sqsubseteq f_c\, v_c}
  {\{\pair[1]{a_c}{R_c}{a_r}{R_r}(v_c)\}\ \mathit{call}_r \lesssim \mathit{call}_c\
   \{\pair[1]{a_c}{R_c}{a_r}{R_r}(f_c\, v_c)\}}
\]
%
The upper premises of the rule are
the specifications of the two calls:
the C call updates the value of $a_c$ from $v$ to $f_c\, v$,
and the Rust call updates the value of $a_r$ from $v$ to $f_r\, v$.
The last premise requires the two updates to preserve refinement:
if $v_r$ refines $v_c$, then $f_r\, v_r$ refines $f_c\, v_c$.
The conclusion is expressed as a relational triple~\cite{relsep07}
$\{P\}\ t \lesssim s\ \{Q\}$, which means that, starting from $P$,
$t$ refines $s$, and $Q$ holds when both terminate;
it states that the value of the pair changes from $v_c$ to $f_c\, v_c$.
The rule is proved by unfolding the representation pair,
applying each specification to its representation predicate,
and folding the results using the last premise.


For example, we can reason about the corresponding \C{sort} calls in \C{has\_duplicate}
using Rule \textsc{Pair-Update}.
Before these calls, we have
$\pair[1]{(\reg{vla}, \reg{n})}{\Arr{i32}}{\reg{s}}{\Vect{i32}}(\mathit{vs})$.
The Rust \C{sort} borrows the buffer of the \C{Vec},
described by $\mathsf{Array}$,
so the rule applies to the C array and this buffer.
By their specifications, both calls update $\mathit{vs}$ to a sorted
permutation, so
both $f_c$ and $f_r$ are $\mathit{sort}$---a function that returns the sorted list;
the values of this pair are lists of defined integers,
for which refinement is equality, so the last premise holds.
After the calls, we thus have the same pair
with value $\mathit{sort}\ \mathit{vs}$.


%% file: sections/5-1-chain.tex
\section{Proving Refinement}
\label{sec:proof}

We now show how to prove the refinement stated in \cref{def:correctness} using
the resources and library specifications introduced in \cref{sec:common}.
We split this refinement into a chain of refinement relations, where only one
relation is program-specific, while the others can be proved once in advance and
applied to every translation (\cref{sec:chain}), and prove the program-specific
part through simulation (\cref{sec:simulation}).


\subsection{Chain of Refinements}
\label{sec:chain}

To explain our refinement chain, we first introduce imaginary operations and
\emph{wrappers} in CRIS.
To establish simulation, we need to reason about memory using resources, and
CRIS enables this with imaginary operations that interact with resources through
events during execution.
Two imaginary operations are important in our discussion:
$\imop{Guarantee}(P)$ provides resources satisfying $P$, and $\imop{Assume}(P)$
receives them;
if $\imop{Assume}$ cannot receive the required resources, it triggers
$\mathsf{UB}$, and if $\imop{Guarantee}$ cannot provide them, it triggers
$\mathsf{NB}$.

CRIS streamlines the introduction of appropriate imaginary operations into
functions using wrappers~\cite{ccr23, cris26}: $\Wfun$ and $\Wcall$ insert these
operations according to the specification of each function.
$\Wfun(M)$ inserts $\imop{Take}$ and $\imop{Assume}$ for the precondition at the start of each
function in $M$ and $\mathsf{Choose}$ and $\imop{Guarantee}$ for the postcondition at its end.
$\Wcall(M)$ inserts, around each function call in $M$, $\mathsf{Choose}$ and $\imop{Guarantee}$ for
the callee's precondition before the call and $\imop{Take}$ and $\imop{Assume}$ for the
postcondition after it.
For example, $\Wfun(\mathit{lib_C})$ defines \C{sort} as follows:
\[\small
\begin{array}{@{}l@{\ }l@{}}
\Wfun(\mathit{lib_C})(\C{sort}) = \lambda v.
  & \mathit{vs} \leftarrow \mathsf{trigger}\ (\imop{Take}\ (\mathit{list}\ \mathbb{Z}));\
    \mathsf{trigger}\ (\imop{Assume}\ P_{\mathit{sort}}(\mathit{vs})); \\
  & r \leftarrow \mathit{lib_C}(\C{sort})(v); \\
  & r' \leftarrow \mathsf{trigger}\ (\mathsf{Choose}\ \mathit{Val});\
    \mathsf{trigger}\ (\imop{Guarantee}\ Q_{\mathit{sort}}(\mathit{vs}, r, r'));\
    \mathsf{Ret}\ r'
\end{array}
\]
where $P_{\mathit{sort}}$ and $Q_{\mathit{sort}}$ are the pre- and postconditions
of \C{sort} in \cref{fig:sort-specs}.
If a function has no specification, $\Wfun$ and $\Wcall$ insert no imaginary
operations into the function's body or around calls to it, respectively.



Now, the refinement stated in \cref{def:correctness} can be proved using the
following chain of refinements:
\[\small
\begin{array}{@{}l@{\;}l@{\qquad}l@{}}
  & \mathit{program}(F_{Rust},\Wfun(\mathit{lib_{Rust}})) & \\
  = & \mathit{main}(F_{Rust}) \star \Wfun(\mathit{lib_{Rust}}) \star \mathit{lib_{LLVM}} \\
  \ctxref & \Hide(\mathit{main(F_{Rust})},\ \Wfun(\mathit{lib_{Rust}}) \star \mathit{lib_{LLVM}})
  & \text{(1) inlining} \\
  \ctxref & \Hide(\Wfun(\Wcall(\mathit{main(F_C)})),\ \Wfun(\mathit{lib_C}) \star \mathit{lib_{LLVM}})
  & \text{(2) simulation (\cref{sec:simulation})} \\
  \behref & \Wfun(\Wcall(\mathit{main(F_C)})) \star \Wfun(\mathit{lib_C}) \star \mathit{lib_{LLVM}}
  & \text{(3) uninlining} \\
  \behref & \mathit{main(F_C)} \star \mathit{lib_C} \star \mathit{lib_{LLVM}}
  & \text{(4) cancellation} \\
  = & \mathit{program}(F_C,\mathit{lib_C})
\end{array}
\]

%


\vspace{-\parskip}
Step 2 is the core of the chain:
it shows that the Rust program refines the C program through simulation, as
detailed in \cref{sec:simulation}.
Since wrappers are applied to modules in this step, we can interact with
resources through the inserted imaginary operations during simulation.
Here, $\mathit{main(F_C)}$ is wrapped, but $\mathit{main(F_{Rust})}$ is not,
because simulation in CRIS provides initial resources to the source (the C side)
but not to the target (the Rust side) of the refinement.
In addition, library function calls are inlined in these modules:
for modules $M$ and $L$, the module $\Hide(M, L)$ keeps only functions from $M$,
replacing any $\mathsf{Call}$ event to a function from $L$ with the body of that
function.
Inlining is required because C and Rust library functions may differ in their
signatures, preventing corresponding $\mathsf{Call}$ events from being matched
in simulation.




Steps 1 and 3 perform inlining and uninlining, and we prove these refinement
relations in the lemma below:

\begin{lemma}[Inlining and Uninlining]
\label{lem:transform}
For modules $M$ and $L$, the following hold:
\begin{enumerate}[label=(\roman*)]
\item (Inlining)
  $M \star L \ctxref \Hide(M, L)$.
\item (Uninlining)
  If $M$ defines an entry and $L$ does not,%
  \footnote{Since CRIS modules may also maintain their own state and spawn threads,
  this further requires that $L$ maintains no state and no function spawns a function in $L$.}
  then $\Hide(M, L) \behref M \star L$.
\end{enumerate}
\end{lemma}

\begin{proof}[Proof sketch]
{\it(i)}\, Calls to $M$'s functions behave the same in both modules;
if a context calls a function in $L$,
the call triggers $\mathsf{UB}$ in $\Hide(M, L)$, which allows every behavior.
{\it(ii)}\, Execution starts from the entry of $M$,
so $L$'s functions are called only from within those of $M$.
$\Hide(M, L)$ runs them in place, as $\sem{M \star L}$ does at
each call, so every behavior of $\Hide(M, L)$ is a behavior of $M \star L$.
\end{proof}


Finally, step 4 removes wrappers using the cancellation theorem of CRIS, which
states that $\Wfun(\Wcall(M))$ $\behref$ $M$ if $M$ defines an entry and the
precondition of the entry holds at the start.
This holds because $\imop{Guarantee}$ of a caller, inserted by $\Wcall$, matches
$\imop{Assume}$ of the callee, inserted by $\Wfun$, and vice versa.
In step 4, the precondition of the entry requires the authorities of the two
initial memories, which can be established at the start of the program, so we
can apply the cancellation theorem.
Note that $\Wfun(\Wcall(\mathit{main(F_C)})) \star \Wfun(\mathit{lib_C}) \star
\mathit{lib_{LLVM}}$ is the same as $\Wfun(\Wcall(\mathit{main(F_C)} \star
\mathit{lib_C} \star \mathit{lib_{LLVM}}))$:
$\mathit{lib_C}$ has no function calls, so $\Wcall$ does not introduce any
imaginary operations;
we provide no specifications for $\mathit{lib_{LLVM}}$, so wrappers insert no
imaginary operations;
and wrappers distribute over linking.


%% file: sections/5-2-simulation.tex
\subsection{Establishing Simulation}
\label{sec:simulation}

We now show how to prove step 2 through simulation.
The adequacy theorem of CRIS states that for functions with the same name in two
modules, if the execution of the target function can be matched by that of the
source function with the same $\mathsf{IO}$ events, then the target contextually
refines the source.

The beginning and end of the simulation are handled by the lemmas stated in
\cref{sec:ca}.
\cref{lem:registration} provides a representation pair for the input before the
LLVM IR functions are called, and, by \cref{lem:observation}, if the two
functions maintain the representation pair, we obtain refinement between the
outputs.




The remaining task is to prove that the ITrees denoted by the two LLVM IR
functions $F_{\mathit{C}}$ and $F_{\mathit{Rust}}$ maintain the representation
pair and relate their return values, as described by the following relational
triple:
\[\small
\begin{array}{@{}l@{}}
\{\mathsf{Auth}(m_c) \ast \mathsf{Auth}(m_r) \ast \pair[1]{a_c}{R_c}{a_r}{R_r}(v)\} \\
\quad \sem{F_{\mathit{Rust}}}(\pf{args}_{\mathit{Rust}}(v, a_r), m_r)
  \lesssim \sem{F_{\mathit{C}}}(\pf{args}_{\mathit{C}}(v, a_c), m_c) \\
\{\lambda (r_r, m_r')\, (r_c, m_c').\
  \mathsf{Auth}(m_c') \ast \mathsf{Auth}(m_r') \ast
  \exists v'.\ \pair[\dq]{a_c}{R_c}{a_r}{R_r}(v') \ast \pure{r_r \sqsubseteq r_c}\}
\end{array}
\]
The precondition consists of the memory authorities and the resources from
\cref{lem:registration};
the postcondition consists of the authorities of the final memories, the
resources required by \cref{lem:observation}, and refinement between the return
values.
We call this triple the \emph{obligation}, as it is the only part proved
separately for each translation.

\begin{figure}[t]
\[
\footnotesize
\begin{array}{@{}c@{}}
s, s', t, t' : \mathsf{itree}\ E\ A
\qquad s_0 : \mathsf{itree}\ E\ B
\qquad k : B \to \mathsf{itree}\ E\ A
\qquad \mathit{reduces} \subseteq \mathsf{itree}\ E\ A \times \mathsf{itree}\ E\ A
\\[8pt]
  s \Downarrow_{Y} \text{ if every execution of } s \text{ ends with } \mathsf{Ret}
  \text{ without triggering any event in } Y;
  \mathsf{IO} \text{ denotes the set of } \mathsf{IO} \text{ events}
\\[8pt]
\inferrule*{ }{\mathit{reduces}(s, s)}
\qquad
\inferrule*{s_0 \Downarrow_{E} \\ s_0 \text{ ends with } \mathsf{Ret}\ v \\ \mathit{reduces}(k\ v, s')}
  {\mathit{reduces}((x \leftarrow s_0;\ k\ x),\ s')}
\\[8pt]
\inferrule*[right=\textsc{Reduce-C}]
  {\mathit{reduces}(s, s') \quad \{P\}\ t \lesssim s'\ \{S\}}
  {\{P\}\ t \lesssim s\ \{S\}}
\qquad
\inferrule*[right=\textsc{Reduce-Rust}]
  {\mathit{reduces}(t, t') \quad \{P\}\ t' \lesssim s\ \{S\}}
  {\{P\}\ t \lesssim s\ \{S\}}
\\[8pt]
\inferrule*[right=\textsc{Call-C}]
  {\mathit{lib_C}(f)(a) \Downarrow_{\mathsf{IO}} \quad
   \{P\}\ f(a)\ \{\lambda r.\ Q(r)\} \\\\
   \forall r.\ \{Q(r) \ast R\}\ t \lesssim k\ r\ \{S\}}
  {\{P \ast R\}\ t \lesssim (r \leftarrow f(a);\ k\ r)\ \{S\}}
\quad
\inferrule*[right=\textsc{Call-Rust}]
  {\mathit{lib_{Rust}}(f)(a) \Downarrow_{\mathsf{IO}} \quad
   \{P\}\ f(a)\ \{\lambda r.\ Q(r)\} \\\\
   \forall r.\ \{Q(r) \ast R\}\ k\ r \lesssim s\ \{S\}}
  {\{P \ast R\}\ (r \leftarrow f(a);\ k\ r) \lesssim s\ \{S\}}
\end{array}
\]
\caption{Representative simulation rules derived from the rules of CRIS.}
\label{fig:sim-rules}
\end{figure}

%

To prove the obligation, we use the simulation rules provided by CRIS, as well
as a few convenience rules we derived from CRIS's rules, with representative
ones shown in \cref{fig:sim-rules}.
Rules \textsc{Reduce-C} and \textsc{Reduce-Rust} execute code that triggers no
events on one side;
$\mathit{reduces}(s, s')$ states that $s$ can be reduced to $s'$ by executing a
part of $s$ that triggers no events.
Computations on registers and memory that do not cause UB trigger no events,
so these rules can be applied to them.
In contrast, calls to functions with specifications cannot be reduced,
because the wrappers introduce imaginary operations around them,
and we introduce separate rules for handling these calls:
Rules \textsc{Call-C} and \textsc{Call-Rust} execute a library call using the
specification of the callee.
The rules provide the precondition $P$ and continue, for every return value $r$,
with the postcondition $Q(r)$, preserving the other resources $R$.
Both rules also require the library function to return without triggering $\mathsf{IO}$ events.
Note that these rules are usable even though we inline functions because ITrees
are coinductively defined and their unfolding is lazy;
a function is inlined only when a term with a $\mathsf{Call}$ event is unfolded,
and we can apply these rules before the unfolding.
Since these two rules execute a program on one side,
the two programs need not make corresponding calls.
When two calls that return without $\mathsf{IO}$ events correspond, Rule \textsc{Pair-Update} stated in \cref{sec:spec}
can be used.
It is derived from the above two rules:
it unfolds the pair, applies \textsc{Call-C} and \textsc{Call-Rust},
and folds the results back.

\begin{figure}[t]
\centering
\scriptsize
\newcommand{\ind}{\hspace*{0.6em}}
\newcommand{\irc}[1]{\textcolor{lstcomment}{\texttt{\itshape ;\ #1}}}
\resizebox{\ifdim\width>\linewidth\linewidth\else\width\fi}{!}{%
\begin{tabular}{@{}c@{\hspace{0.5em}}l@{\enspace}l@{\enspace}l@{}}
 & \textbf{Rust}
  & \textbf{Resources / Trace}
  & \textbf{C} \\
\midrule
\cnum{a} & $\mathit{vs} \leftarrow \mathsf{trigger}\ \mathsf{IO}(\texttt{"input"}, ())$
  & $\mathsf{Interact}\ (\texttt{"input"}, (), \mathit{vs})$
  & $\mathit{vs} \leftarrow \mathsf{trigger}\ \mathsf{IO}(\texttt{"input"}, ())$ \\
 & $(a_r, m_r) \leftarrow \pf{register}_{\mathit{Rust}}(\mathit{vs}, m_0)$
  & $\res{\{\mathsf{pair}^{a}(\mathit{vs})\}}$
  & $(a_c, m_c) \leftarrow \pf{register}_{\mathit{C}}(\mathit{vs}, m_0)$ \\
 & \irsk{define} \irst{i1} \irsf{@has\_duplicate}\texttt{(}\irst{ptr} \texttt{\%a.0, }\irst{i64} \texttt{\%a.1)}
  & $\res{\{\mathsf{pair}^{a}(\mathit{vs})\}}$
  & \irsk{define} \irst{i1} \irsf{@has\_duplicate}\texttt{(}\irst{ptr} \texttt{\%a, }\irst{i64} \texttt{\%n)} \\
\midrule
\cnum{b} & \ind\texttt{\%s = }\irsk{alloca} \irst{[24 x i8]}
  & $\res{\{\mathsf{pair}^{a}(\mathit{vs})\}}$
  & \ind\texttt{\%vla = }\irsk{alloca} \irst{i32}\texttt{, }\irst{i64} \texttt{\%n} \\
 &
  & $\res{\{\mathsf{pair}^{a}(\mathit{vs})\}}$
  & \ind\irc{\%0 = 4 * \%n} \\
 & \ind\irsk{call} \irsf{@to\_vec}\texttt{(\%s, \%a.0, \%a.1)}
  & $\res{\{\mathsf{pair}^{a}(\mathit{vs}) \ast \mathsf{pair}^{s}(\mathit{vs})\}}$
  & \ind\irsk{call} \irsf{@llvm.memcpy}\texttt{(\%vla, \%a, \%0)} \\
\midrule
\cnum{c} & \ind\irc{\%0 = \%s + 8, \%1 = \%s + 16}
  & $\res{\{\mathsf{pair}^{a}(\mathit{vs}) \ast \mathsf{pair}^{s}(\mathit{vs})\}}$
  & \ind\irc{\%add.ptr = \%vla + 4 * \%n} \\
 & \ind\texttt{\%buf = }\irsk{load} \irst{ptr}\texttt{, }\irst{ptr} \texttt{\%0}
  & $\res{\{\mathsf{pair}^{a}(\mathit{vs}) \ast \mathsf{pair}^{s}(\mathit{vs})\}}$
  & \\
 & \ind\texttt{\%len = }\irsk{load} \irst{i64}\texttt{, }\irst{ptr} \texttt{\%1}
  & $\res{\{\mathsf{pair}^{a}(\mathit{vs}) \ast \mathsf{pair}^{s}(\mathit{vs})\}}$
  & \\
 & \ind\irsk{call} \irsf{@sort}\texttt{(\%buf, \%len)}
  & $\res{\{\mathsf{pair}^{a}(\mathit{vs}) \ast \mathsf{pair}^{s}(\mathit{ws})\}}$
  & \ind\irsk{call} \irsf{@sort}\texttt{(\%vla, \%add.ptr)} \\
\midrule
\cnum{d} & \texttt{loop}\ $\rightarrow$ \texttt{body}, \texttt{exit}
  & $\res{\{\mathsf{pair}^{a}(\mathit{vs}) \ast \mathsf{pair}^{s}(\mathit{ws}) \ast I_{\mathit{loop}}\}}$
  & \texttt{for.body}\ $\rightarrow$ \texttt{for.body}, \texttt{cleanup} \\
 & \texttt{body}\ $\rightarrow$ \texttt{loop}, \texttt{exit}
  &
  & \\
\midrule
\cnum{e} & \texttt{exit:}
  & $\res{\{\mathsf{pair}^{a}(\mathit{vs}) \ast \mathsf{pair}^{s}(\mathit{ws})\}}$
  & \texttt{cleanup:} \\
 & \ind\irsk{call} \irsf{@drop\_in\_place}\texttt{(\%s)}
  & $\res{\{\mathsf{pair}^{a}(\mathit{vs}) \ast \pure{\reg{r} = \reg{more}}\}}$
  & \ind\irc{\%r = phi(false, \%cmp2)} \\
 & \ind\irsk{ret} \irst{i1} \texttt{\%more}
  & $\res{\{\mathsf{pair}^{a}(\mathit{vs}) \ast \pure{\reg{r} = \reg{more}}\}}$
  & \ind\irsk{ret} \irst{i1} \texttt{\%r} \\
\midrule
\cnum{f} & $o_r \leftarrow \mathit{observe}(r_r)$
  & $\res{\{\mathsf{pair}^{a}(\mathit{vs}) \ast \pure{o_r = o_c}\}}$
  & $o_c \leftarrow \mathit{observe}(r_c)$ \\
 & $o'_r \leftarrow \mathit{observe}(\pf{read}_{\mathit{Rust}}(a_r, m_r'))$
  & $\res{\{\mathsf{pair}^{a}(\mathit{vs}) \ast \pure{o_r = o_c \land o'_r = o'_c}\}}$
  & $o'_c \leftarrow \mathit{observe}(\pf{read}_{\mathit{C}}(a_c, m_c'))$ \\
 & $\mathsf{Ret}\ (\mathsf{Some}\ (o_r, o'_r))$
  & $\mathsf{Done}\ (\mathsf{Some}\ (o_c, o'_c))$
  & $\mathsf{Ret}\ (\mathsf{Some}\ (o_c, o'_c))$
\end{tabular}%
}
\caption{Proof outline of \texttt{has\_duplicate}.
The left and right columns show corresponding code of the two drivers
and of the two IRs in \cref{fig:example},
where comments summarize the instructions that only compute register values;
the middle column shows the resources after each row,
omitting the memory authorities and the ownership of local allocations,
or the trace element that both sides produce.}
\label{fig:proof-outline}
\end{figure}

Now, we demonstrate simulation using these rules with a proof of
\C{has\_duplicate}, whose outline is shown in \cref{fig:proof-outline}.
The proof involves two representation pairs:
$\mathsf{pair}^{a}$ denotes the input, $\pair[1]{(\reg{a},
\reg{n})}{\Arr{i32}}{(\reg{a.0}, \reg{a.1})}{\Arr{i32}}$,
and $\mathsf{pair}^{s}$ denotes the local copies, $\pair[1]{(\reg{vla},
\reg{n})}{\Arr{i32}}{\reg{s}}{\Vect{i32}}$.
We split the proof into six parts, \cnum{a}--\cnum{f}, where
\cref{lem:registration} handles \cnum{a} and \cref{lem:observation} handles
\cnum{f}.


In \cnum{b}, Rules \textsc{Reduce-C} and \textsc{Reduce-Rust}
execute \C{alloca}, which provides ownership of the local copies;
\textsc{Reduce-C} also executes \C{memcpy}, which is an LLVM intrinsic with
no specification, and
\textsc{Call-Rust} applies the specification of \C{to\_vec}.
The two copies then represent $\mathit{vs}$,
so their predicates fold into $\mathsf{pair}^{s}(\mathit{vs})$.
If the VLA exceeds the stack allocation limit,
the C \C{alloca} triggers $\mathsf{UB}$ and the simulation ends there,
so the proof need not exclude such inputs.

In \cnum{c}, the two \C{sort} calls update $\mathsf{pair}^{s}(\mathit{vs})$ to $\mathsf{pair}^{s}(\mathit{ws})$,
where $\mathit{ws}$ is the sorted permutation of $\mathit{vs}$.
Since the Rust \C{sort} is called on the buffer of the \C{Vec},
which \textsc{Reduce-Rust} loads,
the proof applies \textsc{Pair-Update} to the C array and this buffer,
unfolding and folding the predicates as follows:
\[\footnotesize
\begin{array}{@{}l@{\qquad}l@{}}
\{\mathsf{pair}^{s}(\mathit{vs})\} & \\
\{\rep[1]{(\reg{vla}, \reg{n})}{(\Arr{i32})}{\mathit{vs}} \ast \rep[1]{\reg{s}}{(\Vect{i32})}{\mathit{vs}}\}
  & \text{unfold } \mathsf{pair}^{s} \\
\{\rep[1]{(\reg{vla}, \reg{n})}{(\Arr{i32})}{\mathit{vs}} \ast \reg{s} \mapsto_1 (c, \reg{buf}, \reg{len}) \\
\quad \ast\ \rep[1]{(\reg{buf}, \reg{len})}{(\Arr{i32})}{\mathit{vs}}\}
  & \text{unfold } \mathsf{Vec} \\
\{\reg{s} \mapsto_1 (c, \reg{buf}, \reg{len}) \ast \pair[1]{(\reg{vla}, \reg{n})}{\Arr{i32}}{(\reg{buf}, \reg{len})}{\Arr{i32}}(\mathit{vs})\}
  & \text{fold the pair} \\
\quad \mathtt{@sort}(\reg{buf}, \reg{len}) \lesssim \mathtt{@sort}(\reg{vla}, \reg{add.ptr})
  & \textsc{Pair-Update} \\
\{\reg{s} \mapsto_1 (c, \reg{buf}, \reg{len}) \ast \pair[1]{(\reg{vla}, \reg{n})}{\Arr{i32}}{(\reg{buf}, \reg{len})}{\Arr{i32}}(\mathit{ws})\} & \\
\{\rep[1]{(\reg{vla}, \reg{n})}{(\Arr{i32})}{\mathit{ws}} \ast \rep[1]{\reg{s}}{(\Vect{i32})}{\mathit{ws}}\}
  & \text{unfold the pair, fold } \mathsf{Vec} \\
\{\mathsf{pair}^{s}(\mathit{ws})\}
  & \text{fold } \mathsf{pair}^{s}
\end{array}
\]

\vspace{-\parskip}
In \cnum{d}, the two loops are related using the coinduction
principle~\cite{paco13, gpaco20} provided by CRIS.
The proof chooses an invariant that holds at the start of every iteration of both loops.
It then executes one iteration of each loop under the invariant:
either both loops exit, or both start the next iteration with the invariant holding again.
In the latter case, since both loops have made progress,
coinduction allows the proof to assume that the two loops are related.
Here, the invariant is $\mathsf{pair}^{a}(\mathit{vs}) \ast \mathsf{pair}^{s}(\mathit{ws}) \ast I_{\mathit{loop}}$,
where $I_{\mathit{loop}}$ states that the C index is $i$ and the Rust window starts at $i - 1$
for some $1 \le i < n$.
In each iteration, both loops compare $\mathit{ws}[i-1]$ and $\mathit{ws}[i]$;
if they are equal or $i + 1 = n$, both loops exit with the same result,
and otherwise both start the next iteration with $i + 1$, where the invariant holds again.

In \cnum{e}, \textsc{Call-Rust} applies the specification of \C{drop\_in\_place},
which frees the buffer of the \C{Vec},
and the two functions release their stack allocations upon return, including the local copies.
Both return the same value, and the resources at return,
$\mathsf{pair}^{a}(\mathit{vs}) \ast \pure{\reg{r} = \reg{more}}$,
satisfy the postcondition of the obligation.



%% file: sections/6-automation.tex
\section{Automating Translation Validation with AI Agents}
\label{sec:automation}

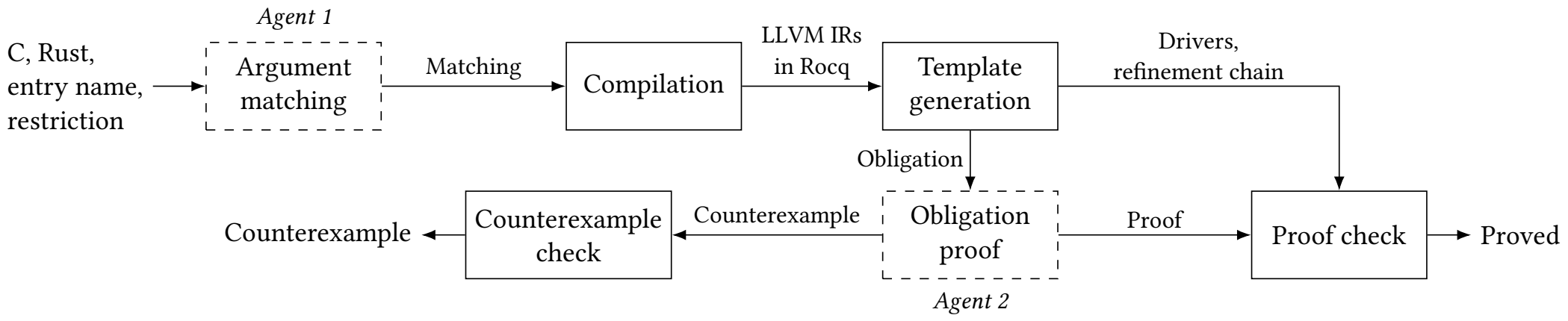
\begin{figure}[t]
\centering
\small
\resizebox{\ifdim\width>\linewidth\linewidth\else\width\fi}{!}{%
\begin{tikzpicture}[
  >={Latex[length=1.6mm]},
  stage/.style={draw, thin, inner sep=3pt, minimum height=10mm, minimum width=20mm, align=center},
  agent/.style={stage, dashed},
  role/.style={font=\footnotesize\itshape, inner sep=1pt},
  part/.style={font=\footnotesize, inner sep=2pt, align=center},
]
\node[align=left] (in) at (0,0) {C, Rust,\\ entry name,\\ restriction};
\node[agent, right=6mm of in] (match) {Argument\\ matching};
\node[role, above=1mm of match] {Agent 1};
\node[stage, right=21mm of match] (compile) {Compilation};
\node[stage, right=16mm of compile] (gen) {Template\\ generation};
\draw[->] (in) -- (match);
\draw[->] (match) -- node[part, above] {Matching} (compile);
\draw[->] (compile) -- node[part, above] {LLVM IRs\\ in Rocq} (gen);
\node[agent] (proof) at (gen |- 0,-1.7) {Obligation\\ proof};
\node[role, below=1mm of proof] {Agent 2};
\node[stage, right=22mm of proof] (check) {Proof check};
\node[right=5mm of check] (proved) {Proved};
\node[stage, left=24mm of proof] (judge) {Counterexample\\ check};
\node[left=5mm of judge] (ctx) {Counterexample};
\draw[->] (gen) -- node[part, left] {Obligation} (proof);
\draw[->] (gen.east) -| node[part, above, pos=0.25] {Drivers,\\ refinement chain}
  (check.north);
\draw[->] (proof) -- node[part, above] {Proof} (check);
\draw[->] (proof) -- node[part, above] {Counterexample} (judge);
\draw[->] (check) -- (proved);
\draw[->] (judge) -- (ctx);
\end{tikzpicture}%
}
\caption{The translation validation workflow.
Dashed boxes are agentic steps; solid boxes are algorithmic steps;
labels show what each step passes on.}
\label{fig:workflow}
\end{figure}

We fully automate the translation validation process using AI agents in the
workflow depicted in \cref{fig:workflow}.
It involves two agentic steps: argument matching and obligation proof.
All other steps are implemented as algorithmic procedures without involving
agents.
In this section, we discuss each step of the workflow.


\paragraph{\textbf{Input.}}

The input includes a C source, its Rust translation, the name of the public
entry function to be validated, and a restriction on the function's input space.
The current implementation provides a list of predefined restrictions, and the
user may select one of them for each argument:
an integer is nonzero/nonnegative/positive; a list is nonempty; a list contains
only nonnegative/positive elements; a string contains only
binary-digit/decimal-digit/lowercase characters.
Extending the implementation to support an arbitrary Rocq predicate would not be
challenging.


\paragraph{\textbf{Argument matching.}}

Agent~1 determines the correspondence between the arguments of the two functions
and assigns a common abstraction to each pair.
For example, for \C{has\_duplicate}, the agent matches the C arguments \C{(a,
n)} with the Rust argument \C{a} and assigns List as their common abstraction.
A function may have multiple argument pairs with common abstractions, e.g., when
taking two lists as input.


\paragraph{\textbf{Compilation.}}

We use the same compiler settings for every input (see \cref{app:compilation}
for the exact settings).
We disable optimizations that may cause the two LLVM IRs to differ unnecessarily
in structure;
for example, we disable loop unrolling since the two compilers may unroll a loop
by different factors.
We also prevent inlining of each library function call if the callee
is defined in our library modules, and let the compilers inline the other library calls;
in \C{has\_duplicate}, \C{windows} and \C{any} are inlined to
form a loop as in C, whereas the \C{sort} functions in both C and Rust are not
inlined and can be reasoned about using their specifications.


After compilation, since our LLVM IR semantics does not support \C{undef},
we replace \C{undef} in the C-side LLVM IR with \C{poison}
and \C{undef} in the Rust-side LLVM IR with zero or \C{null}, depending on the value's type.
This conversion may add behaviors to the C program
and remove behaviors from the Rust program, making the refinement easier to establish.
However, the validation remains sound in most cases,
assuming that the translation uses only safe Rust.
In the C-side IR, \C{undef} is either never read or represents the value of
an uninitialized variable, whose read is already UB in the C source.
In the Rust-side IR, \C{undef} is never read if the compiler is correct,
since safe Rust never reads a value before initializing it.

\paragraph{\textbf{Template generation.}}

We generate a template for the refinement proof, consisting of the two drivers
and the refinement chain, where the obligation of the simulation proof is the
only part left to be proved by an agent in a subsequent step.
Generating drivers requires defining $\pf{valid\_input}_F$,
$\pf{register}_{F,X}$, $\pf{args}_{F,X}$, and $\pf{read}_{F,X}$.
Their exact behavior depends on the argument types and the argument matching.
For each function, we prepare a building block for each supported argument type
and common abstraction listed in \cref{tab:common-abstractions} and compose them
to complete the drivers even when multiple argument pairs are found.
In addition, $\pf{valid\_input}_F$ combines the given restriction with the
condition that the input fits within the size bounds of the architecture.


\paragraph{\textbf{Obligation proof.}}

Agent~2 proves the given obligation using the provided simulation rules.
The proof ideas described in \cref{sec:simulation} are provided as skills to
guide the agent.
The agent can only modify its proof files and is not allowed to weaken the final
theorem stating the refinement.


\paragraph{\textbf{Proof check.}}

Once the agent completes the proof, the Rocq kernel checks it, and we also ensure
that no additional axioms or admits have been introduced. A failed proof is
returned to the agent for revision.

%

\paragraph{\textbf{Counterexample check.}}

When the agent determines that the refinement is unprovable, it proposes a
counterexample, which is an input on which the two LLVM IR programs behave
differently.
To validate the counterexample, we run the two programs on the suggested input.
If the C-side program incurs UB, the counterexample is invalid even when the
outputs differ between C and Rust.
For this reason, we extract the ITree of the C-side driver into
OCaml~\cite{itree19, vellvm21} and execute it, detecting any $\mathsf{UB}$ event
and obtaining results that may contain \C{poison}.
We then run the Rust program natively and compare its results with those of C up
to value refinement.
An invalid counterexample is returned to the agent.


\paragraph{\textbf{Output.}}

The workflow ends with one of three outcomes: proved, counterexample, or fail,
where fail covers every outcome other than proved or counterexample.
A counterexample does not necessarily mean that the translation is incorrect
because validation is performed at the LLVM IR level.
In particular, if the C source has UB but it becomes defined behavior in the
LLVM IR, we may obtain a false counterexample.
Therefore, the user should review the counterexample to determine whether it is
genuine.


%% file: sections/7-1-eval.tex
\section{Evaluation}
\label{sec:evaluation}

In this section, we describe our implementation of \toolname{}
(\cref{sec:implementation}), evaluate it on 519 C programs (\cref{sec:vert}),
and conduct a case study on a larger C program (\cref{sec:qoi}).
Our experiments were conducted on an Ubuntu machine with an AMD Ryzen 9 9950X
16-core processor (16 cores, 32 threads, 4.3 GHz) and 128 GB of DRAM.

\subsection{Implementation}
\label{sec:implementation}

\toolname{} is implemented in Rocq~9.0 on top of the CRIS~\cite{cris26} and
ITrees~\cite{itree19} libraries.
%
Our development adds no axioms:
the eight axioms it relies on are all assumed by the libraries it uses.
We use Codex CLI and Claude Code to run agents.
To improve proof construction, we set the context window of Agent~2 to 872k
tokens in Codex CLI.
Our library modules include 29 LLVM intrinsics, 25 C library functions, and 49
Rust library functions.
These include C functions for memory allocation, strings, math, characters, and
I/O, and Rust functions for \C{Vec}, slices, \C{str}, memory allocation, I/O, and
panics.
For Rust \C{Vec} and slices, the supported element types are \C{i32}, \C{u32},
\C{i64}, \C{u64}, \C{u8}, \C{bool}, and \C{char}, as well as nested \C{Vec} for
two-dimensional arrays.
Our trusted computing base comprises the Rocq kernel, the definitions of ITrees
and CRIS modules, from which behaviors and refinement are defined, the Vellvm
parser that turns LLVM IR files into Rocq definitions, our Vellvm-based LLVM IR
semantics, including its memory model, and our library modules.


\subsection{Evaluation on the VERT Benchmarks}
\label{sec:vert}

We evaluate \toolname{} on the benchmark suite of VERT~\cite{vert25}, which
includes 519 C programs.%
\footnote{The VERT dataset contains 520 C programs, but we exclude two obviously
incorrect programs that lack \texttt{return} statements in some branches and split
one program with two public functions into two programs, yielding 519 programs
in total.}
Of these, 504 programs are from the TransCoder-IR dataset~\cite{transcoderir23},
implementing algorithms involving loops and recursion over integers,
floating-point numbers, arrays, strings, and two-dimensional arrays.
The other 15 programs are parts of projects collected from prior work on
C-to-Rust translation~\cite{emre2021translating, crown23}, and they use structs
with pointers.
Each program has 13.3 LOC on average, with a maximum of 154 LOC.
We translated each C program into Rust using \C{gpt-6-sol} via Codex CLI, with a
prompt requesting a safe, idiomatic translation.
We manually determined the input restrictions for each program.

%

We evaluate \toolname{} with the following three research questions (RQs):
\begin{enumerate}[label=\textbf{RQ\arabic*.}, leftmargin=*]
\item \textit{Effectiveness}: For how many translations does \toolname{}
  prove correctness or find a counterexample, and
  how do the results compare with existing C-to-Rust translation validation tools?
  (\cref{sec:rq1})
\item \textit{Efficiency}: How much time and cost does translation validation require?
  (\cref{sec:rq2})
\item \textit{Impact of model choice}: How do effectiveness and efficiency
  vary with the agent model?
  (\cref{sec:rq3})
\end{enumerate}
For RQ1 and RQ2, we used \C{gpt-6-luna} as Agent~1 and \C{gpt-6.1-sol}
as Agent~2, both via Codex CLI.

\subsubsection{RQ1: Effectiveness}
\label{sec:rq1}

Of the 519 translations, \toolname{} proves the correctness of 486, finds a
counterexample for 26, and fails on 7, demonstrating the effectiveness of our
approach.
The seven failures are due to unsupported features: four involve recursive
structs, one involves an \C{Option<String>} return value, one involves the
\C{mustprogress} LLVM IR attribute, which our LLVM IR semantics does not model, and
one involves closures.





\begin{figure}[t]
\newsavebox{\cexAc}\newsavebox{\cexAr}\newsavebox{\cexBc}\newsavebox{\cexBr}%
\newsavebox{\cexCc}\newsavebox{\cexCr}\newsavebox{\cexDc}\newsavebox{\cexDr}%
\newlength{\cexh}%
\newcommand{\cexcell}[3]{\setlength{\cexh}{\dp#2}\ifdim\dp#3>\cexh\setlength{\cexh}{\dp#3}\fi
  \vbox to \cexh{\hbox{#1}\vfil}}%
\newcommand{\cexlabel}[1]{%
  \begingroup
  \setcaptionsubtype\phantomcaption\label{#1}%
  \makebox[2.2em][l]{\scriptsize\smash{\raisebox{-\ht\strutbox}{\footnotesize\thesubfigure}}}%
  \endgroup}%
\lstset{aboveskip=0pt, belowskip=0pt}%
\begin{lrbox}{\cexAc}\begin{minipage}[t]{\dimexpr0.4\textwidth-2.2em}
\begin{lstlisting}[style=c, emph={strlen}]
// s: char[]
int n = strlen(s);
int arr[n + 1];
for (int i = 1; i <= n; i++) {
  arr[i] = s[i - 1] - '0';
}
\end{lstlisting}
\end{minipage}\end{lrbox}%
\begin{lrbox}{\cexAr}\begin{minipage}[t]{0.57\textwidth}
\begin{lstlisting}[style=rust, emph={len, chars, enumerate}]
// s: &str
let n = s.len() as i32;
let mut arr = vec![0_i32; (n + 1) as usize];
for (i, ch) in s.chars().enumerate() {
  arr[i + 1] = ch as i32 - '0' as i32;
}
\end{lstlisting}
\end{minipage}\end{lrbox}%
\begin{lrbox}{\cexBc}\begin{minipage}[t]{\dimexpr0.4\textwidth-2.2em}
\begin{lstlisting}[style=c, emph={calloc}]
// n: size_t, buf->data: char *
buf->len = n;
buf->data = calloc(n + 1, 1);
return buf;
\end{lstlisting}
\end{minipage}\end{lrbox}%
\begin{lrbox}{\cexBr}\begin{minipage}[t]{0.57\textwidth}
\begin{lstlisting}[style=rust, emph={wrapping_add, vec}]
// n: usize, buf.data: Vec<u8>
buf.len = n;
buf.data = vec![0; n.wrapping_add(1)];
return buf;
\end{lstlisting}
\end{minipage}\end{lrbox}%
\begin{lrbox}{\cexCc}\begin{minipage}[t]{\dimexpr0.4\textwidth-2.2em}
\begin{lstlisting}[style=c]
// n, sum: int
bool subset[n + 1][sum + 1];
\end{lstlisting}
\end{minipage}\end{lrbox}%
\begin{lrbox}{\cexCr}\begin{minipage}[t]{0.57\textwidth}
\begin{lstlisting}[style=rust, emph={vec}]
// n, sum: i32
let subset =
  vec![vec![false; (sum + 1) as usize]; (n + 1) as usize];
\end{lstlisting}
\end{minipage}\end{lrbox}%
\begin{lrbox}{\cexDc}\begin{minipage}[t]{\dimexpr0.4\textwidth-2.2em}
\begin{lstlisting}[style=llvm]
%n1 = add nsw i32 %n, 1
%rows = zext i32 %n1 to i64
%s = add i32 %sum, 1
%cols = zext i32 %s to i64
%size = mul nuw i64 %cols, %rows
%subset = alloca i8, i64 %size
\end{lstlisting}
\end{minipage}\end{lrbox}%
\begin{lrbox}{\cexDr}\begin{minipage}[t]{0.57\textwidth}
\begin{lstlisting}[style=llvm, emph={@from_elem}]
%row = alloca [24 x i8]
%s = add i32 %sum, 1
%cols = sext i32 %s to i64
call void @from_elem(ptr %row, i1 false, i64 %cols)
\end{lstlisting}
\end{minipage}\end{lrbox}%
\begin{minipage}[t]{0.4\textwidth}
\cexcell{\cexlabel{fig:counterexamples:a}\usebox{\cexAc}}{\cexAc}{\cexAr}
\smallskip\hrule\smallskip
\cexcell{\cexlabel{fig:counterexamples:b}\usebox{\cexBc}}{\cexBc}{\cexBr}
\smallskip\hrule\smallskip
\cexcell{\cexlabel{fig:counterexamples:c}\usebox{\cexCc}}{\cexCc}{\cexCr}
\smallskip\hrule\smallskip
\cexcell{\cexlabel{fig:counterexamples:d}\usebox{\cexDc}}{\cexDc}{\cexDr}
\end{minipage}\hfill\vrule\hfill
\begin{minipage}[t]{0.57\textwidth}
\cexcell{\usebox{\cexAr}}{\cexAc}{\cexAr}
\smallskip\hrule\smallskip
\cexcell{\usebox{\cexBr}}{\cexBc}{\cexBr}
\smallskip\hrule\smallskip
\cexcell{\usebox{\cexCr}}{\cexCc}{\cexCr}
\smallskip\hrule\smallskip
\cexcell{\usebox{\cexDr}}{\cexDc}{\cexDr}
\end{minipage}
\caption{C code (left) and Rust code (right) of translations
for which a counterexample is found:
(a)--(b) incorrect translations,
(c) a translation whose counterexample
causes UB,
and (d) the LLVM IRs compiled from (c).}
\label{fig:counterexamples}
\end{figure}

We manually investigated the 26 counterexamples and found that 3 are genuine
counterexamples, while the other 23 are false positives.
We now discuss representative cases in detail.

\cref{fig:counterexamples:a} shows one incorrect translation, which fails to
preserve the behavior on large inputs.
The C program stores the length of a given string in \C{int n}, truncating the
64-bit result of \C{strlen}.
For example, if a string has length $2^{32}+3$, \C{n} is 3, and C copies only
the first 3 characters into \C{arr} of length 4.
The Rust program also truncates the length, but it iterates over the entire
string and panics due to an out-of-bounds index for \C{arr}.
This example demonstrates the effectiveness of \toolname{} in finding subtle
translation errors.

The other two incorrect translations share the same cause, and we show one case
in \cref{fig:counterexamples:b}.
The C program calls \C{calloc}, which returns \C{NULL} when allocation fails,
but the Rust program uses \C{vec!}, which panics on allocation failure.
Strictly speaking, this does not preserve the behavior of C, but one may want to
consider such a translation correct, as \C{vec!} is an idiomatic way of
allocating memory in Rust, and allocation failure is rare in practice.
For this purpose, we could modify the models of the allocation functions to
never fail.

The 23 false positives occur when the C program exhibits UB that is compiled
into defined behavior in LLVM IR.
Of these, 20 arise from signed integer overflow and 3 from accesses beyond a row
of a two-dimensional array.
\cref{fig:counterexamples:c} shows one case:
when \C{sum} equals \C{INT\_MAX}, \C{sum + 1} overflows, which is UB in C.
However, the C-side LLVM IR in \cref{fig:counterexamples:d} does not result in
UB:
\C{sum + 1} is not marked as non-wrapping, so it wraps around to \C{INT\_MIN};
\C{zext} zero-extends it to $2^{31}$, and an array of this size is allocated.
In contrast, the Rust-side LLVM IR sign-extends \C{INT\_MIN} to a huge size with
\C{sext}, and \C{from\_elem} panics due to a capacity overflow.



We also manually reviewed the compiled LLVM IR programs to confirm that our
removal of \C{undef} does not affect these outcomes.
In these programs, all occurrences of \C{undef} are operands of \C{phi} or
\C{select}:
in 9 C programs and 62 Rust programs, \C{undef} appears but is never read;
in 5 C programs, it represents the value of an uninitialized variable, and
reading it is already UB in C.

\paragraph{\textbf{Comparison with existing tools.}}

We compare the effectiveness of \toolname{} with that of two existing C-to-Rust
translation validation tools, VERT~\cite{vert25} and
RustAssure~\cite{rustassure25}, using the same translations.


VERT compares each translation with
a Rust program obtained by compiling the C program to WebAssembly and then
translating it to Rust using rWasm~\cite{bosamiya2022provably}.
It first runs property-based testing and then performs bounded model checking,
each for 120 seconds, unrolling each loop up to 14 iterations.
We classify its outcome as \emph{validated} if both testing
and bounded model checking pass,
\emph{counterexample} if either check finds a difference,
or \emph{fail} if it produces no result, e.g., due to a timeout or
a compilation or runtime error.

RustAssure symbolically executes the LLVM IR of the two programs
and compares graphs representing their symbolic return values using edit distance.
It limits symbolic execution to 180 minutes per function
with input arrays of 100 elements.
We classify its outcome as \emph{validated} if it finds no difference,
\emph{difference} if it reports a difference,
or \emph{fail} if it produces no result.
A reported difference only indicates a potential difference in the output and
does not provide an input that can serve as a counterexample.

Note that a validated outcome from either tool does not provide a correctness
guarantee as strong as that of \toolname{}.
Their validation covers only a subset of possible inputs and handles loops only
up to a bound, failing to cover all possible executions.
In contrast, \toolname{} considers all valid inputs with respect to
user-provided input restrictions and executions of arbitrary length, including
nonterminating ones, using coinduction.



\begin{table}[t]
\small
\caption{Results of \toolname, VERT, and RustAssure on the 519 translations.}
\label{tab:rq1}
\begin{tabular}{@{}lrrrrrrr@{}}
\toprule
\multicolumn{1}{c}{\multirow{2}{*}{\toolname}}
  & \multicolumn{3}{c}{VERT} & \multicolumn{3}{c}{RustAssure} & \multirow{2}{*}{\bf Total} \\
\cmidrule(lr){2-4} \cmidrule(l){5-7}
  & Validated
  & Counterexample
  & Fail
  & Validated
  & Difference
  & Fail \\
\midrule
  Proved & 152 & 120 & 214 & 142 & 68 & 276 & {\bf 486} \\
  Counterexample & 5 & 12 & 9 & 5 & 6 & 15 & {\bf 26} \\
  Fail & 1 & 0 & 6 & 0 & 0 & 7 & {\bf 7} \\
\midrule
  {\bf Total} & {\bf 158} & {\bf 132} & {\bf 229} & {\bf 147} & {\bf 74} & {\bf
  298} & {\bf 519} \\
\bottomrule
\end{tabular}
\end{table}

The experimental results, summarized in \cref{tab:rq1}, show that \toolname{} is
more effective than the existing tools.
Of the 519 translations, VERT validates 158, finds a counterexample for 132, and
fails on 229;
RustAssure validates 147, reports a difference for 74, and fails on 298.
We now discuss the differences in the results.



VERT reports a counterexample for 120 translations that \toolname{} verifies,
but none of them is a genuine counterexample:
65 are due to providing invalid inputs to functions, e.g., passing an array
pointer and a symbolic integer independent of the array length, when the
function expects the correct length, resulting in out-of-bounds access;
32 are due to checking strict equality of outputs rather than refinement, e.g.,
rejecting a case where the C program has UB and the Rust program panics;
23 are due to incorrect preprocessing that changes the behavior before the
checks, e.g., replacing \C{f64} with \C{f32}.

RustAssure reports a difference for 68 translations that \toolname{} verifies;
it only provides different graphs, not a counterexample.
We investigated how different graphs arise from symbolic execution and
identified two notable causes:
it sometimes fails to match C and Rust arguments, providing different symbolic
inputs to the two programs;
when symbolic execution times out, it compares incomplete graphs.


For the three incorrect translations discovered by \toolname{}, neither tool
provides a valid counterexample.
For \cref{fig:counterexamples:a}, VERT provides a counterexample consisting of
an invalid input, rather than a genuine one involving an extremely long string;
RustAssure only reports a difference.
Both tools fail on the two cases involving structs with heap buffers.
For the 23 false positive cases from \toolname{}, VERT and RustAssure each
validate 5 cases.
However, these results arise only because they cover a limited set of inputs,
not because they appropriately handle UB in C at the source level.
VERT samples inputs randomly and checks only small values during bounded model
checking, while RustAssure explores only some paths within its time limit and
fixes some inputs to constants.
Neither tool exercises extreme input values such as \C{INT\_MAX}.


%

\begin{figure}[t]
\includegraphics{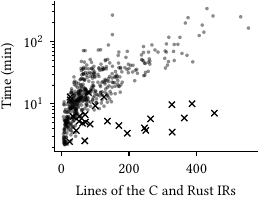}\hfill
\includegraphics{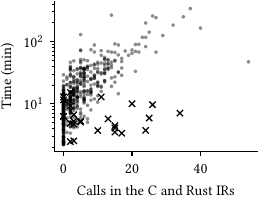}\hfill
\includegraphics{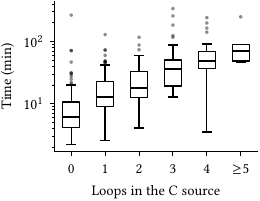}\\[1em]
\includegraphics{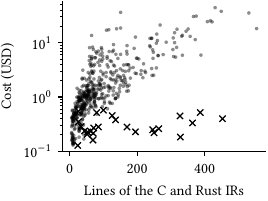}\hfill
\includegraphics{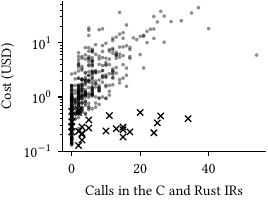}\hfill
\includegraphics{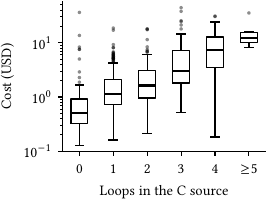}
\caption{Time (top) and cost (bottom) of validating the 512 translations
by the lines and the calls of the C and Rust IRs and the loops of the C source;
in the scatter plots, dots are proved translations and crosses are counterexamples,
and the whiskers of the box plots extend to 1.5 times the interquartile range.}
\label{fig:efficiency}
\end{figure}

\subsubsection{RQ2: Efficiency}
\label{sec:rq2}

We assess the efficiency of \toolname{} by measuring the time and cost of
validation.
We measured the end-to-end runtime of the entire workflow and counted the tokens
used, calculating the cost based on the public API pricing of each model.
On average, across the 512 translations excluding the 7 failures, \toolname{}
took 23.3 minutes and cost \$2.91, with Agent~2 accounting for almost all of the
time (98.1\%) and cost (99.9\%).
For the 486 proved cases, it took 24.2 minutes and cost \$3.05;
for the 26 cases with counterexamples, it took 6.6 minutes and cost \$0.31.
Most cases took much less than the average, with medians of 13.0 minutes and
\$1.09.
The distributions are skewed toward a few translations, with 90th percentiles of
46.6 minutes and \$6.84.
The most expensive case took 338.4 minutes and cost \$44.31 and contains a
triple-nested loop.

%
%

To better understand how program characteristics affect time and cost, we
investigate how they correlate with the number of lines and the number of calls
in the LLVM IRs from C and Rust, and the number of loops in the C source,
showing the results in \cref{fig:efficiency}.
These features correspond to the work involved in our proofs:
lines represent instructions that the proof simulates, calls are where it
applies library specifications and matches the calls of the two programs, and
loops are where it requires coinduction and invariants.
We measure the correlations using Spearman's rank correlation coefficient
$\rho$, which measures how consistently one value increases with the other:
it is 1 when their rankings coincide and 0 when there is no rank correlation.
For time and cost, $\rho$ is 0.76 and 0.73 for lines, 0.67 and 0.69 for calls,
and 0.63 and 0.59 for loops (all $p < 0.001$), so the correlations are strongest
for lines, followed by calls and loops.



\subsubsection{RQ3: Impact of model choice}
\label{sec:rq3}

To see how the model affects the effectiveness and efficiency of \toolname{}, we
measured the time and cost of running Agent~2 with four models on 21 programs.
We selected seven programs from each of three difficulty levels defined by the
time \C{gpt-6.1-sol} spent on them in RQ2: at most 10 minutes (easy), 30 to 60
minutes (medium), and at least 140 minutes (hard).
We compare three OpenAI models, \C{gpt-6.1-terra}, \C{gpt-6.1-sol}, and
\C{gpt-6-astra}, in increasing order of general performance and API price, using
Codex CLI, and Anthropic's frontier model, \C{claude-opus-5-5}, using Claude
Code.


\begin{table}[t]
\small
\caption{Proved translations, average time (minutes), and average cost (USD) of Agent~2
with four models on the 21 programs, seven at each level of difficulty.}
\label{tab:rq3}
\begin{tabular}{@{}lrrrrrrr@{}}
\toprule
& & \multicolumn{2}{c}{Easy (7)} & \multicolumn{2}{c}{Medium (7)} & \multicolumn{2}{c}{Hard (7)} \\
\cmidrule(lr){3-4} \cmidrule(lr){5-6} \cmidrule(l){7-8}
Model & Proved & Time & Cost & Time & Cost & Time & Cost \\
\midrule
\texttt{gpt-6.1-terra} & 20 & 10.8 & 0.65 & 85.3 & 12.67 & 200.9 & 40.77 \\
\texttt{gpt-6.1-sol} & 21 & 7.7 & 0.57 & 41.5 & 5.73 & 193.4 & 27.29 \\
\texttt{gpt-6-astra} & 21 & 8.8 & 3.86 & 46.4 & 27.80 & 150.4 & 155.65 \\
\texttt{claude-opus-5-5} & 21 & 6.8 & 2.87 & 52.9 & 27.64 & 136.6 & 60.78 \\
\bottomrule
\end{tabular}
\end{table}

The experimental results, summarized in \cref{tab:rq3}, show that our approach
generalizes to other models.
Only \C{gpt-6.1-terra} failed to finish one hard program within the six-hour
time limit;
all other models completed all programs.
Therefore, we believe that \toolname{} can use most of today's frontier models
for running agents, while less capable models may not be suitable for validating
difficult translations.

In addition, the time and cost vary across models, so the choice of model can
affect the efficiency of our approach.
In terms of time, \C{gpt-6-astra} and \C{claude-opus-5-5} do not show meaningful
improvements over \C{gpt-6.1-sol} for easy and medium programs, and their
advantages are clear only for hard programs.
In terms of cost, \C{gpt-6.1-sol} finished at the lowest cost across all levels,
although \C{gpt-6.1-terra} has the lowest per-token price.
This shows that a cheaper model does not necessarily lead to a lower validation
cost, and one needs to find a balance between the model's capability and its
price to reduce the validation cost.


%% file: sections/7-2-eval.tex
\subsection{Case Study: QOI}
\label{sec:qoi}

Beyond the VERT benchmarks, we conduct a case study on a larger program to
demonstrate how \toolname{} handles more complex features.
We use QOI,\footnote{\url{https://qoiformat.org/}} a lossless image compression
format~\cite{qoi22}, whose reference implementation in C consists of 312 lines
and four functions:
\C{qoi\_decode} takes a buffer containing compressed data and allocates and
returns a new buffer containing decompressed pixels;
\C{qoi\_read} reads a file, decodes the data with \C{qoi\_decode}, and returns
the pixel buffer;
\C{qoi\_encode} and \C{qoi\_write} perform the inverse operations.
We treat \C{qoi\_read} and \C{qoi\_write} as public functions and validate the
translation of each.
\toolname{} successfully validated \C{qoi\_read} and \C{qoi\_write}, with
Agent~2 using \C{gpt-6.1-sol} taking 12.6 and 7.0 hours, respectively.
We now discuss the validation of \C{qoi\_read} in more detail.


%

%

One important aspect of \C{qoi\_read} is that it returns a pointer to an array.
Its signatures are as follows:

\begin{lstlisting}[style=c]
void *qoi_read(const char *filename, qoi_desc *desc, int channels); // C
\end{lstlisting}
\vspace{-1em}
\begin{lstlisting}[style=rust]
fn qoi_read(filename: &str, desc: &mut QoiDesc, channels: u8) -> Option<Vec<u8>>; // Rust
\end{lstlisting}

\noindent
To compare the return values appropriately as abstract values, Agent~1 also
determines the common abstraction for return values during argument matching,
assigning List in this case.
Furthermore, when this Rust function is compiled into LLVM IR, the return type
becomes \C{void}, and the result is produced by writing it to the memory pointed
to by the first argument \C{\%\_0}:




\begin{lstlisting}[style=llvm]
define void @qoi_read(ptr sret([24 x i8]) %_0, ptr %filename.0, i64 %filename.1, ptr %desc, i8 %channels)
\end{lstlisting}

\noindent
Template generation recognizes this pattern by examining the return type and the
number of parameters and generates the following driver so that the return value
can be observed correctly:
\[\footnotesize
\begin{array}{@{}l@{}}
  \mathit{driver}(\texttt{qoi\_read}_{\mathit{Rust}}) \triangleq \lambda\_.\\
  \quad v \leftarrow \mathsf{trigger}\ \mathsf{IO}(\texttt{"input"}, ());\\
  \quad \mathbf{if}\ \neg\,\pf{valid\_input}_{\texttt{qoi\_read}}(v)\ \mathbf{then}\ \mathsf{Ret}\ \mathsf{None}\\
  \quad \mathbf{else}\ (p, q, m) \leftarrow \pf{register}_{\texttt{qoi\_read},\mathit{Rust}}(v, m_0);\\
  \quad \phantom{\mathbf{else}\ } (\_, m') \leftarrow \sem{\texttt{qoi\_read}_{\mathit{Rust}}}(\pf{args}_{\texttt{qoi\_read},\mathit{Rust}}(v, p, q), m);\\
  \quad \phantom{\mathbf{else}\ } o \leftarrow \mathit{observe}(\pf{read}_{\texttt{qoi\_read},\mathit{Rust}}(q, m'));\\
  \quad \phantom{\mathbf{else}\ } o' \leftarrow \mathit{observe}(\pf{read}_{\texttt{qoi\_read},\mathit{Rust}}(p, m'));\\
  \quad \phantom{\mathbf{else}\ } \mathsf{Ret}\ (\mathsf{Some}\ (o, o'))
\end{array}
\]
\noindent
Its $\pf{register}$ additionally allocates $q$, a memory location where the
return value will be written, and $\pf{args}$ constructs the first argument
using $q$. After the function call, the nominal return value of type \C{void} is
ignored, and the actual return value is read from memory as a list using
$\pf{read}$ with $q$.

Another important aspect is that \C{qoi\_read} performs file operations.
In C, it opens a file with \C{fopen} to obtain a \C{FILE *} handle, reads
through it using \C{fread}, and closes it using \C{fclose};
in Rust, it uses \C{File::open} to obtain a \C{File} object, reads from it using
\C{read\_exact}, and drops it to close the file.
We introduce representation predicates for C \C{FILE} and Rust \C{File} as
follows:
\[ \small
\begin{array}{@{}r@{\;}c@{\;}l@{}}
\rep[\dq]{p}{\mathsf{FILE}}{s} & \triangleq &
  \exists\, \mathit{meta}.\ p \mapsto_{\dq} (s, \mathit{meta}) \\
\rep[\dq]{p}{\mathsf{File}}{s} & \triangleq & p \mapsto_{\dq} s
\end{array}
\]
where $s$ is a 32-bit integer that serves as a file descriptor.
C \C{FILE} consists of a file descriptor and metadata, such as error and
end-of-file indicators and the current position;
Rust \C{File} just wraps the file descriptor.
The common abstraction File relates them with the following representation pair:
\[ \small
\begin{array}{@{}r@{\;}c@{\;}l@{}}
\pair[\dq]{p_c}{\mathsf{FILE}}{p_r}{\mathsf{File}}(s) & \triangleq &
  \rep[\dq]{p_c}{\mathsf{FILE}}{s} \ast \rep[\dq]{p_r}{\mathsf{File}}{s}
\end{array}
\]
The library modules describe I/O functions by triggering distinct $\mathsf{IO}$
events for each kind of operation, consistently on both the C and Rust sides:
\C{fopen} and \C{File::open} trigger an event named \C{file.open};
\C{fread} and \C{read\_exact} trigger \C{file.read};
\C{fclose} and dropping a \C{File} trigger \C{file.close}.
This allows the same events to be matched during simulation.

Note that we do not claim to fully support I/O operations in this work;
the purpose of this case study is to demonstrate that \toolname{} can reason
about I/O by modeling library functions with events in ITrees.
The current implementation includes only a few functions and excludes several
features, such as the buffering behavior of C \C{FILE}.
Furthermore, events alone would not be sufficient:
for example, a translation may read a file through a single call, while the
source reads the same data through multiple calls, resulting in a different
number of events and causing simulation to fail.
Validating translations that heavily involve I/O would be an interesting
direction for future work.

%% file: sections/8-related.tex
\section{Related Work}
\label{sec:related}

\paragraph{\textbf{C-to-Rust translation.}}
C2Rust~\cite{c2rust} syntactically translates C into unsafe Rust,
and other rule-based translators aim to produce safer Rust,
either by improving the output of C2Rust or by translating C directly:
some replace raw pointers with references or other smart pointers
by analyzing ownership and aliasing~\cite{emre2021translating, emre2023aliasing, crown23,
fromherz2026scylla, chen2026refinator},
others replace C types and APIs with their Rust counterparts,
such as tagged unions, generics, algebraic data types, locks, and I/O~\cite{hong2024tag,
kim2026securing, wu2025genc2rust, hong2024dont, hong2026replacing, hong2023concrat, hong2025forcrat},
and techniques for different features can be combined~\cite{proctor26}.
LLM-based translators combine LLMs with program analysis
or iterative repair~\cite{hong2024type, xu2025optimizing, zhang2025systematic,
cai2026rustmap, yuan2026project, shiraishi2026smartc2rust, zhou2025c2rusttv}.
Most of these studies evaluate the correctness of their translations through testing.


\paragraph{\textbf{Translation validation.}}
Pnueli et al.~\cite{pnueli98} proposed translation validation,
which checks each translation instead of the translator.
It has been applied to optimizing compilers~\cite{necula2000translation, barrett2005tvoc},
and its validators have themselves been verified
in the verified compiler CompCert~\cite{compcert09, tristan2008formal, tristan2009verified}.
LLVM IR has since become a major target of translation validation:
Alive2~\cite{alive221} validates LLVM optimizations with SMT queries over bounded executions,
and Crellvm~\cite{crellvm18} lets the optimizations generate proofs that a verified checker validates.
These validators relate programs in the same language,
whereas we relate IR programs compiled from two different languages.
Translation validation has also been applied to C-to-binary compilation~\cite{sewell2013translation},
optimized x86 binaries~\cite{sharma2013data},
and V8's JIT compiler~\cite{kwon2024translation}.

\paragraph{\textbf{Refinement theory.}}
Our proofs build on CRIS~\cite{cris26}, a Rocq framework for contextual refinement
that generalizes Conditional Contextual Refinement~\cite{ccr23}, which unifies refinement and separation logic.
They follow a line of work on proving contextual refinement
using relational separation logics~\cite{relsep07, simuliris22};
their resources follow separation logic~\cite{seplogic01} as formalized in Iris~\cite{iris15, iris18, ipm17},
and their proofs of loops use parameterized coinduction~\cite{paco13, gpaco20}.
Programs in CRIS are ITrees~\cite{itree19},
which represent effectful programs as coinductive trees of events,
so semantics defined using ITrees are both executable and composable by interpretation~\cite{yoon22formal}.

\paragraph{\textbf{LLM-based proof generation.}}
Proofs generated by LLMs still provide formal guarantees
once they are checked by a proof assistant, such as Rocq or Lean, or a verifier.
This idea has increasingly been applied to proving existing theorems in a proof assistant~\cite{baldur23},
verifying programs against their specifications~\cite{autoverus25},
and proving equivalence between a program and its LLM translation~\cite{llmlift24}.
Recently, an LLM proved the memory safety of binaries of real-world libraries
against mechanically generated specifications~\cite{anthropic2026formal}.
We follow this idea and apply it to C-to-Rust translation validation.

\paragraph{\textbf{Formal foundations.}}
Formal semantics of C and Rust have been developed for their own purposes:
for C, to verify a compiler~\cite{compcert09} or to clarify its standards~\cite{cerberus16},
and for Rust, to prove the soundness of its type system~\cite{rustbelt18} or to verify programs against their specifications~\cite{refinedrust24, verus23}.
However, none of them defines the behaviors of C and Rust programs in the same terms,
as required for C-to-Rust translation validation.
We therefore turned to LLVM IR and chose Vellvm~\cite{vellvm21}, a semantics of LLVM IR defined using ITrees.
Another option is to define semantics for C and Rust separately but in the same terms:
translation validation requires semantics for both languages but not a single semantics for both,
just as CompCert~\cite{compcert09} gives each of its languages its own semantics
and relates them through their behaviors.
We leave this option, which would validate translations at the source level, for future work.

%% file: sections/9-conclusion.tex
\section{Conclusion}

We present \toolname{}, a tool that validates C-to-Rust translation by proving
in Rocq that the Rust program refines the C program at the LLVM IR level.
The use of LLVM IR allows us to describe the behavior of programs in the two
languages in the same terms.
We relate the different data representations of C and Rust through common
abstractions and reason about library function calls using their specifications
over abstract values.
Refinement is established through a chain of refinement relations, where
the program-specific part is proved by AI agents.

We evaluate \toolname{} on 519 C programs from VERT.
It effectively validates translations, proving the translations of 486 programs
correct without bounding their executions and finding three incorrect
translations with valid counterexamples.
On average, it validated each translation in 23.3 minutes at a cost of \$2.91
using \C{gpt-6.1-sol}.
We also conduct a case study on QOI, demonstrating how \toolname{} can handle
file operations.

%% file: sections/A-semantics.tex
\section{Semantics of LLVM IR}
\label{app:semantics}

This appendix describes the semantics of LLVM IR.
A subset of the syntax of LLVM IR is shown in \cref{fig:llvm-syntax}.
Metavariables $\mathtt{\%}x$, $\mathtt{@}g$, $\mathtt{@}f$, and $\ell$
range over the names of registers, global variables, functions,
and basic blocks, respectively.
Each instruction in a basic block is paired with a register
that receives its result, if it has one.
\cref{fig:llvm-semantics} shows the denotations of expressions, instructions,
and terminators.
The pure functions $\pf{\mathsf{gep}}$, $\pf{\mathsf{alloca}}$, $\pf{\mathsf{load}}$, and $\pf{\mathsf{store}}$
return a value of the following type:
\[
\mathit{MemoryResult}\ A ::= \mathsf{Success}\ A \mid \mathsf{Failure}
\]
They return $\mathsf{Failure}$ when the operation is undefined,
such as an access outside an allocated block,
and the denotation then triggers $\mathsf{UB}$.
$\pf{\mathit{op}}$ returns its result in the same form,
returning $\mathsf{Failure}$ on a division by zero or an overflow of signed division.

For a block $B$ entered from the block $\ell'$,
$\sem{B.\mathsf{phis}}(\rho, \ell')$ evaluates every phi node of $B$ in $\rho$ for $\ell'$
and assigns their registers at once.
The code of $B$, denoted by $\sem{B.\mathsf{code}}(\rho, m)$,
runs the instructions of $B$ in order, writing each result to the assigned register;
an instruction without one, such as $\mathsf{store}$, leaves $\rho$ unchanged.
\[\small
\begin{array}{@{}l@{}}
\mathit{run}(\rho, m, \ell', \ell) =
  \rho_0 \leftarrow \sem{B_\ell.\mathsf{phis}}(\rho, \ell');\
  (\rho_1, m_1) \leftarrow \sem{B_\ell.\mathsf{code}}(\rho_0, m);\
  o \leftarrow \sem{B_\ell.\mathsf{term}}(\rho_1, m_1); \\
\quad \mathbf{match}\ o\ \mathbf{with}\
  \mathsf{inl}\ \ell'' \Rightarrow \mathsf{Tau}\ (\mathit{run}(\rho_1, m_1, \ell, \ell''))
  \mid \mathsf{inr}\ v \Rightarrow \mathsf{Ret}\ (v, m_1) \\[3pt]
\sem{F}(\mathit{args}, m) =
  \mathit{run}(\{F.\mathsf{args} := \mathit{args}\}, m, F.\mathsf{entry}, F.\mathsf{entry})
\end{array}
\]
%
Here $B_\ell$ is the basic block of $F$ labeled $\ell$,
and $\mathit{run}$ executes $B_\ell$ entered from the block $\ell'$.
When the terminator jumps to a block $\ell''$,
$\mathit{run}$ continues with $B_{\ell''}$ after a silent step;
as in the iteration combinator of the ITrees library, this step guards the recursive call
and makes $\mathit{run}$ well-defined without changing behaviors.
When the terminator returns a value $v$,
$\mathit{run}$ ends with $v$ and the current memory.
$\sem{F}(\mathit{args}, m)$ assigns the arguments to $F.\mathsf{args}$
and runs $F$ from the entry block with the memory $m$.

\begin{figure}[tbp]
\small
\[
\begin{array}{@{}r@{\;}c@{\;}l@{}}
\tau & ::= & \mathtt{i}N \mid \mathtt{float} \mid \mathtt{double}
  \mid \mathtt{ptr} \mid \mathtt{void} \mid [k \times \tau]
  \qquad N \in \{1, 8, 16, 32, 64\} \quad k \in \mathbb{N} \quad n \in \mathbb{Z} \cup \mathbb{R}\\
e & ::= & \mathtt{\%}x \mid \mathtt{@}g \mid n
  \mid \mathsf{binop}(\mathit{op}, e, e)
  \mid \mathsf{icmp}(c, e, e)
  \mid \mathsf{select}(e, e, e)
  \mid \mathsf{getelementptr}(\tau, e, \mathit{list}\ e) \\
i & ::= & e
  \mid \mathsf{call}(\tau, e, \mathit{list}\ e)
  \mid \mathsf{alloca}(\tau, e)
  \mid \mathsf{load}(\tau, e)
  \mid \mathsf{store}(e, e) \\
t & ::= & \mathsf{ret}(e)
  \mid \mathsf{br}(e, \ell, \ell)
  \mid \mathsf{br}(\ell)
  \mid \mathsf{unreachable} \\
\phi & ::= & \mathsf{phi}(\mathit{list}\ (\ell, e)) \\
B & ::= & \{\mathsf{label} : \ell;\
  \mathsf{phis} : \mathit{list}\ (\mathtt{\%}x, \phi);\
  \mathsf{code} : \mathit{list}\ (\mathit{option}\ \mathtt{\%}x, i);\
  \mathsf{term} : t\} \\
F & ::= & \{\mathsf{name} : \mathtt{@}f;\
  \mathsf{type} : \tau\,(\mathit{list}\ \tau);\
  \mathsf{args} : \mathit{list}\ \mathtt{\%}x; \\
& & \phantom{\{}\mathsf{entry} : \ell;\
  \mathsf{blocks} : \mathit{list}\ B\} \\
P & ::= & \{\mathsf{globals} : \mathit{list}\ (\mathtt{@}g, e);\
  \mathsf{functions} : \mathit{list}\ F\}
\end{array}
\]
\caption{A subset of the syntax of LLVM IR.}
\label{fig:llvm-syntax}
\end{figure}

\begin{figure}[tbp]
\small
\begin{tabular}{@{}p{0.5\linewidth}@{}p{0.5\linewidth}@{}}
$\begin{array}[t]{@{}l@{}}
\sem{e} : \mathit{Env} \times \mathit{Mem} \to \mathsf{itree}\ E\ \mathit{Val} \\[4pt]
\sem{\mathsf{binop}(\mathit{op}, e_1, e_2)}(\rho, m) = \\
\quad v_1 \leftarrow \sem{e_1};\ v_2 \leftarrow \sem{e_2};\
  \mathbf{match}\ \pf{\mathit{op}}(v_1, v_2)\ \mathbf{with} \\
\qquad \mid \mathsf{Success}\ v \Rightarrow \mathsf{Ret}\ v \\
\qquad \mid \mathsf{Failure} \Rightarrow \mathsf{trigger}\ \mathsf{UB} \\[2pt]
\sem{\mathsf{icmp}(c, e_1, e_2)}(\rho, m) = \\
\quad v_1 \leftarrow \sem{e_1};\ v_2 \leftarrow \sem{e_2};\ \mathsf{Ret}\ (\pf{c}(v_1, v_2))
\end{array}$
&
$\begin{array}[t]{@{}l@{}}
\sem{\mathsf{select}(e_1, e_2, e_3)}(\rho, m) = \\
\quad b \leftarrow \sem{e_1};\ v_2 \leftarrow \sem{e_2};\ v_3 \leftarrow \sem{e_3}; \\
\quad \mathbf{if}\ b = \texttt{poison}\ \mathbf{then}\ \mathsf{Ret}\ \texttt{poison} \\
\quad \mathbf{else}\ \mathsf{Ret}\ (\mathbf{if}\ b\ \mathbf{then}\ v_2\ \mathbf{else}\ v_3) \\[2pt]
\sem{\mathsf{getelementptr}(\tau, e, \mathit{es})}(\rho, m) = \\
\quad p \leftarrow \sem{e};\ \mathit{vs} \leftarrow \sem{\mathit{es}};\
  \mathbf{match}\ \pf{\mathsf{gep}}(\tau, p, \mathit{vs}, m)\ \mathbf{with} \\
\qquad \mid \mathsf{Success}\ v \Rightarrow \mathsf{Ret}\ v \\
\qquad \mid \mathsf{Failure} \Rightarrow \mathsf{trigger}\ \mathsf{UB}
\end{array}$
\\
\noalign{\vskip 4pt\hrule height \arrayrulewidth\vskip 4pt}
$\begin{array}[t]{@{}l@{}}
\sem{i} : \mathit{Env} \times \mathit{Mem} \to \mathsf{itree}\ E\ (\mathit{Val} \times \mathit{Mem}) \\[4pt]
\sem{\mathsf{alloca}(\tau, e)}(\rho, m) = \\
\quad n \leftarrow \sem{e};\ \mathbf{if}\ n = \texttt{poison}\ \mathbf{then}\ \mathsf{trigger}\ \mathsf{UB} \\
\quad \mathbf{else}\ \mathbf{match}\ \pf{\mathsf{alloca}}(\tau, n, m)\ \mathbf{with} \\
\qquad \mid \mathsf{Success}\ (m', a) \Rightarrow \mathsf{Ret}\ (a, m') \\
\qquad \mid \mathsf{Failure} \Rightarrow \mathsf{trigger}\ \mathsf{UB} \\[2pt]
\sem{\mathsf{load}(\tau, e)}(\rho, m) = \\
\quad p \leftarrow \sem{e};\ \mathbf{match}\ \pf{\mathsf{load}}(\tau, p, m)\ \mathbf{with} \\
\qquad \mid \mathsf{Success}\ v \Rightarrow \mathsf{Ret}\ (v, m) \\
\qquad \mid \mathsf{Failure} \Rightarrow \mathsf{trigger}\ \mathsf{UB}
\end{array}$
&
$\begin{array}[t]{@{}l@{}}
\sem{e}(\rho, m) = v \leftarrow \sem{e};\ \mathsf{Ret}\ (v, m) \\[4pt]
\sem{\mathsf{store}(e_1, e_2)}(\rho, m) = \\
\quad w \leftarrow \sem{e_1};\ p \leftarrow \sem{e_2}; \\
\quad \mathbf{if}\ p = \texttt{poison}\ \mathbf{then}\ \mathsf{trigger}\ \mathsf{UB} \\
\quad \mathbf{else}\ \mathbf{match}\ \pf{\mathsf{store}}(p, w, m)\ \mathbf{with} \\
\qquad \mid \mathsf{Success}\ m' \Rightarrow \mathsf{Ret}\ ((), m') \\
\qquad \mid \mathsf{Failure} \Rightarrow \mathsf{trigger}\ \mathsf{UB} \\[2pt]
\sem{\mathsf{call}(\tau, e, \mathit{es})}(\rho, m) = \\
\quad \mathit{args} \leftarrow \sem{\mathit{es}};\ p \leftarrow \sem{e}; \\
\quad \mathbf{if}\ p = \texttt{poison}\ \mathbf{then}\ \mathsf{trigger}\ \mathsf{UB} \\
\quad \mathbf{else}\ \mathsf{trigger}\ (\mathsf{Call}(\mathit{fn}(p), (\tau, \mathit{args}, m)))
\end{array}$
\\
\noalign{\vskip 4pt\hrule height \arrayrulewidth\vskip 4pt}
$\begin{array}[t]{@{}l@{}}
\sem{t} : \mathit{Env} \times \mathit{Mem} \to \mathsf{itree}\ E\ (\mathit{Label} + \mathit{Val}) \\[4pt]
\sem{\mathsf{ret}(e)}(\rho, m) = v \leftarrow \sem{e};\ \mathsf{Ret}\ (\mathsf{inr}\ v) \\[2pt]
\sem{\mathsf{unreachable}}(\rho, m) = \mathsf{trigger}\ \mathsf{UB}
\end{array}$
&
$\begin{array}[t]{@{}l@{}}
\sem{\mathsf{br}(e, \ell_1, \ell_2)}(\rho, m) = \\
\quad b \leftarrow \sem{e};\ \mathbf{if}\ b = \texttt{poison}\ \mathbf{then}\ \mathsf{trigger}\ \mathsf{UB} \\
\quad \mathbf{else}\ \mathsf{Ret}\ (\mathsf{inl}\ (\mathbf{if}\ b\ \mathbf{then}\ \ell_1\ \mathbf{else}\ \ell_2)) \\[2pt]
\sem{\mathsf{br}(\ell)}(\rho, m) = \mathsf{Ret}\ (\mathsf{inl}\ \ell)
\end{array}$
\end{tabular}
\caption{The semantics of expressions, instructions, and terminators.
On right-hand sides, $\sem{e}$ stands for $\sem{e}(\rho, m)$. Functions in \pf{blue} are pure functions.}
\label{fig:llvm-semantics}
\end{figure}

%% file: sections/B-compilation.tex
\section{Compilation Settings}
\label{app:compilation}

We compile C programs with Clang~21.1.8 and Rust programs with rustc~1.93.1,
both of which are based on LLVM~21,
to LLVM IR for x86-64 Linux at the optimization level O1 (\C{-O1} and \C{-Copt-level=1}).
We use the GNU11 dialect of C (\C{-std=gnu11}) and the 2021 edition of Rust (\C{-{}-edition=2021}).
For both languages, we use the following settings.
\begin{itemize}
\item \emph{Loop unrolling.}\,
  We disable loop unrolling, since the two compilers may unroll a loop by different factors:
  \C{-fno-unroll-loops} for Clang,
  and for rustc, the LLVM options that set the unrolling thresholds and counts to zero
  and disable runtime unrolling, partial unrolling, peeling, and unroll-and-jam
  (\C{-Cllvm-args=-{}-unroll-threshold=0} and others).
\item \emph{Vectorization.}\,
  Since our semantics does not support vector types,
  we disable loop and SLP vectorization
  (\C{-fno-vectorize}, \C{-fno-slp-vectorize},
  \C{-Cno-vectorize-loops}, and \C{-Cno-vectorize-slp}),
  the vectorization of loop idioms (\C{-disable-loop-idiom-vectorize-all}),
  and vector combining (\C{-disable-vector-combine}),
  and we reject an LLVM IR that still contains vector types or instructions.
\item \emph{Inlining.}\,
  Using the LLVM option \C{-force-attribute},
  we add the \C{noinline} attribute to each library function with a specification
  and leave the other library functions to the optimizer.
  For Rust, we also disable inlining at the MIR level (\C{-Zinline-mir=no}),
  which would otherwise inline library functions before LLVM sees this attribute.
\end{itemize}
For C, we also disable floating-point contraction (\C{-ffp-contract=off}),
which may fuse a multiplication and an addition into one operation,
and keep calls to C library functions, such as \C{isupper} and \C{printf}, as calls to those functions
(\C{-D\_\_NO\_CTYPE} and, if needed, \C{-fno-builtin-}$f$).
For Rust, we make panics abort (\C{-Cpanic=abort}),
so that the LLVM IR has no unwinding code.